\documentclass[12pt]{article}

\usepackage{verbatim,amssymb,amsmath,textcomp}				
\usepackage{amsthm}		
\usepackage{bm}
\usepackage{dsfont}			
\usepackage{multirow}
\usepackage[mathscr]{euscript}
\usepackage{fancyhdr}
\usepackage{enumitem}
\usepackage{graphicx}
\usepackage{geometry}
\usepackage{xcolor}
\usepackage{hyperref}
\usepackage{bbm}
\usepackage[section]{placeins}
\usepackage{cleveref}

\usepackage{tabularx, booktabs}
\newcolumntype{Y}{>{\centering\arraybackslash}X}
\usepackage{algorithm}
\usepackage{algpseudocode}

\usepackage{subcaption}
\usepackage{float}
\usepackage{lineno}

\usepackage{setspace}
\usepackage{tikz}
\usetikzlibrary{arrows}
\usepackage{wrapfig}

\usepackage{csquotes}
\usepackage[style=ext-authoryear, maxcitenames=1, giveninits=true, maxbibnames=99, sortcites=false, natbib, uniquename=init]{biblatex} 
\DeclareNameAlias{author}{family-given}
\DeclareNameAlias{editor}{family-given}

\renewbibmacro{in:}{%
	\ifentrytype{article}
	{}
	{\bibstring{in}%
		\printunit{\intitlepunct}}}
\DeclareFieldFormat[article]{pages}{#1}

\usepackage{lscape}

\newtheorem*{Proof*}{Proof}

\newcommand{\R}{\mathbb{R}}
\newcommand{\E}{\mathbb{E}}
\newcommand{\Var}{\operatorname{Var}}

\newcommand{\diag}{\operatorname{diag}}
\newcommand{\Bern}{\operatorname{Bernoulli}}
\newcommand{\one}{\mathbbm{1}}
\newcommand{\logit}{\operatorname{logit}}
\newcommand{\expit}{\operatorname{expit}}
\newcommand{\Normal}{\mathcal{N}}
\newcommand{\N}{\mathcal{N}}
\newcommand{\bvec}[1]{\bm{#1}}
\newcommand{\given}{\,|\,}
\newcommand{\T}{^{\top}}
\newcommand{\indep}{\perp\!\!\!\perp}
\newcommand{\Had}{\odot}
\newcommand{\new}{\textsc{new}}

\newcommand{\br}{{\mathbf r}}
\newcommand{\bs}{{\mathbf s}}
\newcommand{\bw}{{\mathbf w}}
\newcommand{\bx}{{\mathbf x}}
\newcommand{\by}{{\mathbf y}}
\newcommand{\bz}{{\mathbf z}}
\newcommand{\bA}{{\mathbf A}}
\newcommand{\bB}{{\mathbf B}}
\newcommand{\bC}{{\mathbf C}}
\newcommand{\bD}{{\mathbf D}}
\newcommand{\bG}{{\mathbf G}}
\newcommand{\bI}{{\mathbf I}}
\newcommand{\bM}{{\mathbf M}}

\newcommand{\bV}{{\mathbf V}}
\newcommand{\bW}{{\mathbf W}}
\newcommand{\bX}{{\mathbf X}}
\newcommand{\bbeta}{\bm{\beta}}
\newcommand{\bgamma}{\bm{\gamma}}
\newcommand{\btheta}{\bm{\theta}}
\newcommand{\bmu}{\bm{\mu}}
\newcommand{\bzeta}{\bm{\zeta}}
\newcommand{\bGamma}{\bm{\Gamma}}
\newcommand{\bOmega}{\bm{\Omega}}
\newcommand{\bSigma}{\bm{\Sigma}}
\newcommand{\beps}{\bm{\varepsilon}}
\newcommand{\bLambda}{\bm{\Lambda}}

\renewcommand\footnoterule{\kern-3pt \hrule \textwidth 2in \kern 2.6pt}

\def\boxit#1{\vbox{\hrule\hbox{\vrule\kern6pt \vbox{\kern6pt \textcolor{blue}{#1}\kern6pt}\kern6pt\vrule}\hrule}}

\def\authorfootnote#1{{\let\thefootnote\relax\footnotetext{#1}}}

\allowdisplaybreaks

\begin{document}
	\thispagestyle{empty}
	\baselineskip=28pt

	\begin{center}
		{\LARGE{\bf Mean-field Variational Bayes for Sparse Probit Regression
		}}
	\end{center}
\baselineskip=14pt
\vskip 2mm

\begin{center}
		Augusto Fasano and Giovanni Rebaudo
        \vskip 3mm
        ESOMAS Department, University of Torino and Collegio Carlo Alberto
\end{center}
\begin{refsection}

\bigskip
\begin{center}
{\Large{\bf Abstract}} 
\end{center}
We consider Bayesian variable selection for binary outcomes under a probit link with a spike-and-slab prior on the regression coefficients.
Motivated by the computational challenges encountered by Markov chain Monte Carlo (MCMC) samplers in high-dimensional regimes, we develop a mean-field variational Bayes approximation in which all variational factors admit closed-form updates, and the evidence lower bound is available in closed form.
This, in turn, allows the development of an efficient coordinate ascent variational inference algorithm to find the optimal values of the variational parameters.
The approach produces posterior inclusion probabilities and parameter estimates, enabling interpretable selection and prediction within a single framework.
As shown in both simulated and real data applications, the proposed method successfully identifies the important variables and is orders of magnitude faster than MCMC, while maintaining comparable accuracy.

\baselineskip=14pt
\vskip 8mm


\clearpage\pagebreak\newpage
\pagenumbering{arabic}
\newlength{\gnat}
\setlength{\gnat}{14pt} 
\baselineskip=\gnat

\section{Introduction}
From the seminal work of \citet{albert1993bayesian}, Bayesian binary regression has attracted sustained attention over the past years, with a wide variety of computational strategies for probit and logit links, including sampling \citep{holmes2006bayesian, polson2013bayesian, durante2019conjugate, zens2024ultimate} or variational approximations \citep{jaakkola2000bayesian, consonni2007meanfield, fasano2022scalable, zens2025scalable}.
See also \citet{chopin2017leave} and \citet{anceschi2023bayesian} for reviews addressing possible computational approaches.

In high-dimensional problems, a central inferential task is variable selection, often leveraging sparsity-inducing priors.
This model-based approach produces posterior inclusion probabilities and parameter estimates within a single coherent framework, but places heavy demands on posterior computation: as a consequence, standard Markov chain Monte Carlo (MCMC) can face convergence issues or be prohibitively slow as $p$ grows or when $p \gg n$.
See also \citet{chang2016bayesian, chu2025bayesian} for MCMC strategies and related discussions on variable selection in Bayesian probit models for binary and multicategory outcomes.

In this contribution, in order to overcome the above-mentioned limitations of the sampling approaches, we develop a variational Bayes procedure for sparse probit regression with a spike-and-slab prior based on binary masks \citep{ormerod2017variational}.
To this aim, let $y_{i}\in\{0,1\}$ be the binary response and $\bx_{i}\in\R^{p}$ the $p$-dimensional vector of covariates for units $i=1,\ldots,n$.
For $j=1,\ldots,p$, define the variable $\gamma_{j} \in\{0,1\}$ indicating whether predictor $j$ has an effect on the output ($\gamma_{j}=1$) or not ($\gamma_{j}=0$).
Thus, exploiting the characterization of the probit model in terms of partial observations of latent Gaussian variables \citep{albert1993bayesian}, we write the Bayesian probit model with binary masks as
\begin{equation}
\label{eq:probit_model}
\begin{split}
    y_{i}&=\one\{z_{i}>0\},\qquad z_{i}\mid \bbeta,\bgamma \overset{ind}{\sim} \Normal(\bx_{i}\T \bGamma \bbeta,\,1),\quad i=1,\ldots,n,\\
    \bbeta &\sim \Normal_{p}( \bm{0}, \nu^{2}\bI_{p}),\qquad \gamma_{j} \overset{iid}{\sim} \Bern(\rho),\quad j=1,\ldots,p,
\end{split}
\end{equation}
where $\one\{A\}$ denotes the indicator function of the event $A$, $\bI_{p}$ is the $p$-dimensional identity matrix, $\bbeta=(\beta_{1},\ldots,\beta_{p})\T$, $\bgamma=(\gamma_{1},\ldots,\gamma_{p})\T$, and $\bGamma=\diag(\bgamma)$, while $\nu^{2}>0$ and $\rho\in(0,1)$ represent known hyperparameters.
We thus encode selection through a binary mask in the linear predictor $\bx_{i}\T \bGamma\bbeta$ while keeping $\bbeta\sim\Normal_{p}( \bm{0},\nu^{2} \bI_{p})$ independent of $\bgamma$.
Note that if we set $\tilde{\bbeta} =\bGamma\bbeta$, $\tilde{\beta}_{j} = \gamma_{j}\beta_{j}$ has a mixture prior with a point mass at $0$ with probability $1-\rho$ and $\Normal(0,\nu^{2})$ with probability $\rho$, which coincides with the classical spike-and-slab prior representation for $\tilde{\bbeta}$.

The interest is in inferring which variables truly affect the output and possibly the intensity of such an effect.
Calling $\by=(y_{1},\ldots,y_{n})$, we thus study the posterior inclusion probabilities (PIPs), $\Pr[\gamma_{j}=1\mid \by]$, and the posterior distribution of $\gamma_{j} \beta_{j}$, $j=1,\ldots,p$.
These quantities do not admit closed-form expressions and might be computationally intensive to obtain via MCMC methods.
To overcome these limitations, in Section~\ref{sec2} we propose a mean-field variational Bayes approach, leveraging methods developed by \citet{ormerod2017variational} for linear models.

\section{Mean-field variational Bayes for sparse probit}
\label{sec2}
In the proposed approach, we approximate the joint posterior distribution $p(\bbeta,\bz,\bgamma\mid \by)$ of model~\eqref{eq:probit_model} with the following mean-field variational Bayes (MFVB) approximation
\begin{equation*}
    q(\bbeta,\bz,\bgamma)=q(\bbeta)\;q(\bz)\;\prod_{j=1}^{p} q(\gamma_{j}).
    \nonumber
\end{equation*}
Importantly, the mean-field factorization is imposed across blocks $(\bbeta,\bz,\bgamma)$ and within the $\bgamma$ block, but not within $\bbeta$.
In particular, as shown in the following, $q(\bbeta)$ is a multivariate Gaussian with a full covariance matrix, so dependencies among regression coefficients are allowed, mitigating the Bayesian uncertainty underestimation issues typically associated with fully factorized mean-field approximations that can arise, e.g., in high-dimensional settings or in the presence of collinearity.
Note that in the probit case with a Gaussian prior, \citet{fasano2022scalable} develop a partially factorized variational Bayes approach designed for computational scalability, with theoretical guarantees in high-dimensional settings when $p$ diverges.
However, the computational tractability of their approach relies on the Gaussian prior assumption on the regression coefficients and does not directly extend to spike-and-slab variable selection with latent inclusion indicators, which requires different algorithmic solutions.

The form of the optimal factors is then given by the mean-field equations; see, e.g., \citet{blei2017variational}.
According to those equations, if $\btheta=(\btheta_{1},\ldots,\btheta_{K})$ is the vector of all model parameters, partitioned into $K$ subcomponents $\btheta_{1},\ldots,\btheta_{K}$, and the posterior distribution $p(\btheta\mid\by)$ is approximated by the variational density $q(\btheta)=\prod_{k=1}^K q(\btheta_{k})$, then the optimal variational approximation $q(\btheta_{k})$ satisfies $\log q(\btheta_{k}) = \E_{q(\btheta_{-k})}\left[\log p(\btheta_{k}\mid \btheta_{-k},\by) \right] + \text{const}$, where $\btheta_{-k}$ denotes the vector $\btheta$ with the $k$-th subcomponent $\btheta_{k}$ removed; this notational convention will be used also in the following.
We now derive the form of the optimal $q(\bbeta)$, $q(\bz)$, and $q(\gamma_{j})$, $j=1,\ldots,p$.
We denote by $\phi(t)$ and $\Phi(t)$ the standard normal density and cumulative distribution function, respectively.
Similarly, $\phi_{p}(\bx;\bSigma)$ denotes the density of a $p$-variate Gaussian random variable with mean $\boldsymbol{0}$ and covariance matrix $\bSigma$, evaluated at $\bx$.

\subsection{The optimal variational factor \texorpdfstring{$q(\bbeta)$}{q(beta)}}
\label{subsec:2.1}
From the optimal mean-field equations \citep{blei2017variational}, the optimal $q(\bbeta)$ must satisfy
\begin{equation*}
 \log q(\bbeta)
 = \E_{q(\bz)q(\bgamma)} \big[\log p(\bbeta\mid \bz,\bgamma,\by) \big] + \text{const},
\end{equation*}
where the full conditional $p(\bbeta\mid\bz,\bgamma,\by)= p(\bbeta\mid\bz,\bgamma)$ coincides with the posterior in a Gaussian regression with design $\bX\bGamma$ and unit error variance.
See also \citet{ormerod2017variational}.
One can derive it easily since, in general, $$\log p(\bbeta\mid \bz,\bgamma) = \log p(\bz\mid \bbeta,\bgamma) + \log p(\bbeta) + \text{const}.$$
Calling $\bX=[\bx_{1},\dots,\bx_{n}]\T\in\R^{n\times p}$, for model \eqref{eq:probit_model} it holds
\begin{align*}
 \log p(\bz\mid \bbeta,\bgamma) + \log p(\bbeta)
 &= -\frac{1}{2} \left(\bz - \bX\bGamma\bbeta\right)\T \left(\bz - \bX\bGamma\bbeta\right) - \frac{1}{2\nu^{2}}\,\bbeta\T\bbeta + \text{const},
\end{align*}
where the quadratic term can be expanded as 
\begin{align*}
 \left(\bz - \bX\bGamma\bbeta\right)\T \left(\bz - \bX\bGamma\bbeta\right)
 &= \bz\T \bz - 2\bbeta\T \bGamma \bX\T \bz + \bbeta\T \bGamma \bX\T \bX \bGamma \bbeta\\
 &= \bz\T \bz - 2\bbeta\T \bGamma \bX\T \bz + \bbeta\T(\bGamma \bG\bGamma)\bbeta,
\end{align*}
with $\bG = \bX\T \bX$.
Hence
\begin{align*}
 \log p(\bbeta\mid \bz,\bgamma)
 &= -\frac{1}{2} \bbeta\T\big(\nu^{-2}\bI_{p}+\bGamma \bG\bGamma\big)\bbeta + \bbeta\T \bGamma \bX\T \bz + \text{const},
\end{align*}
i.e., $p(\bbeta\mid \bz,\bgamma)=\phi_{p}(\bbeta -\bV\bGamma \bX\T \bz; \bV)$ with $\bV=(\nu^{-2}\bI_{p}+\bGamma \bG\bGamma)^{-1}$.

Thus, we have
\begin{align*}
 \log q(\bbeta) &= \E_{q(\bz)q(\bgamma)} \big[ - \frac{1}{2}\bbeta\T\big(\nu^{-2}\bI_{p}+\bGamma \bG\bGamma\big)\bbeta + \bbeta\T \bGamma \bX\T \bz \big] + \text{const}\\
 &= -\frac{1}{2} \bbeta\T\big(\nu^{-2}\bI_{p} + \E_{q(\bgamma)}[\bGamma \bG \bGamma]\big)\bbeta + \bbeta\T \E_{q(\bgamma)}[\bGamma] \bX\T \E_{q(\bz)}[\bz] + \text{const},
\end{align*}
from which we can already recognize the $\log$ of a multivariate Gaussian kernel.
In order to write conveniently the parameters, we define $\bar{\bz}=\E_{q(\bz)}[\bz]$ and $\bW=\E_{q(\bgamma)}[\bGamma]=\diag(\bw)$, with $\bw=\E_{q(\bgamma)}[\bgamma]\in [0,1]^{p}$ and $\bOmega = \E_{q(\bgamma)}[\bgamma \bgamma\T]= \bW(\,\bI_{p}-\bW\,)+\bw\bw\T$, so that $\Omega_{jj}=w_{j}=\E_{q(\gamma_{j})}[\gamma_{j}]$ and $\Omega_{jk}=w_{j} w_{k}$ for $j\neq k$.
Exploiting now the fact that
\begin{align*}
  \bGamma \bG \bGamma
  &= (\bX\bGamma)\T (\bX\bGamma) = \begin{bmatrix}
     \gamma_{1} \bX_{:1} & \cdots & \gamma_{p} \bX_{:p}
   \end{bmatrix}\T
   \begin{bmatrix}
     \gamma_{1} \bX_{:1} & \cdots & \gamma_{p} \bX_{:p}
   \end{bmatrix} \\
  &= \big[\,\gamma_{j}\gamma_{k}\, \bX_{:j}\T \bX_{:k}\,\big]_{j,k=1}^{p} = \bG \Had (\bgamma\bgamma\T),
\end{align*}
where $\Had$ denotes the Hadamard (elementwise) product and $\bX_{:j}$ represents the $j$-th column of $\bX$, we have $\E_{q(\bgamma)}[\bGamma \bG \bGamma]=\bG\Had \bOmega$.
Thus, in conclusion, the optimal variational factor $q(\bbeta)$ satisfies
\begin{equation}
\label{eq:betaVB}
q(\bbeta) = \phi_{p}(\bbeta - \bmu; \bSigma),
\end{equation}
where $\bSigma = \big(\nu^{-2}\bI_{p} + \bG \Had \bOmega\big)^{-1}$ and $\bmu = \bSigma\, \bW \bX\T \bar{\bz}$.
Note that since $\bG=\bX\T \bX\succeq 0$ and $\bOmega=\E_{q(\bgamma)}[\bgamma\bgamma\T]\succeq 0$, the Schur product theorem gives $\bG\Had\bOmega\succeq 0$.
Hence $\nu^{-2}\bI_{p}+\bG\Had\bOmega$ is positive definite for any $\nu^{2}>0$.
If some $w_{j}$ degenerate to zero or near-zero values, some simplifications or approximations can be used, avoiding the $\mathcal{O}(p^{3})$ cost that would be required by the direct computation of $\bSigma$ through inversion of the $p\times p$ precision matrix.
See Section~\ref{sec:3} and the Supplementary Material for more details.

\subsection{The optimal variational factor \texorpdfstring{$q(\bz)$}{q(z)}}
\label{subsec:2.2}
Starting with the computation of the full conditional of $\bz=(z_{1},\ldots,z_{n})\T$, notice that
\begin{equation*}
    \begin{split}
        p(\bz\mid\bbeta, \bgamma, \by) &\propto p(\bz\mid \bbeta,\bgamma) p(\by \mid\bz)\\
        &\propto \prod_{i=1}^{n} \phi(z_{i}-\bx_{i}\T\bGamma\bbeta) \one \{z_{i}\in\mathcal{A}_{y_{i}}\},
    \end{split}
\end{equation*}
where $\mathcal{A}_{y_{i}}=(0,\infty)$ if $y_{i}=1$ and $(-\infty,0]$ otherwise.
That is, adapting \citet{albert1993bayesian}, conditionally on $(\bbeta,\bgamma, \by)$ the $z_{i}$ are independently distributed as Gaussian random variables with mean $\bx_{i}\T\bGamma\bbeta$ and unit variance, truncated to the set $\mathcal{A}_{y_{i}}$.
From the optimality equations, for $\bz\in\mathcal{A}_{\by}$, where $\mathcal{A}_{\by} = \mathcal{A}_{y_{1}} \times \cdots \times \mathcal{A}_{y_{n}}$, one gets
\begin{align*}
 \log q(\bz) = \sum_{i=1}^{n} \log q(z_{i}),
\end{align*}
where, for $z_{i}\in\mathcal{A}_{y_{i}}$,
\begin{align*}
    \log q(z_{i}) & = \E_{q(\bbeta)q(\bgamma)}\big[
     \log \phi(z_{i}-\bx_{i}\T\bGamma\bbeta) \big] + \text{const}\\
    &= -\frac{1}{2} \E_{q(\bbeta)q(\bgamma)}\big[ (z_{i} - \bx_{i}\T\bGamma\bbeta)^{2} \big] + \text{const}\\
    &=-\frac{1}{2} z_{i}^{2} + z_{i}\bx_{i}\T \E_{q(\bbeta)q(\bgamma)}\big[ \bGamma\bbeta \big] + \text{const}\\
    &=-\frac{1}{2} z_{i}^{2} + z_{i}\bx_{i}\T \E_{q(\bgamma)}\big[ \bGamma \big] \E_{q(\bbeta)}\big[\bbeta \big] + \text{const}\\
    &=-\frac{1}{2} z_{i}^{2} + z_{i}\bx_{i}\T \bW \bmu + \text{const}\\
    &=-\frac{1}{2} \big(z_{i} - \bx_{i}\T \bW \bmu \big)^{2} + \text{const}.
\end{align*}
Hence, the optimal variational approximation $q(\bz)$ factorizes as $q(\bz)=\prod_{i=1}^{n} q(z_{i})$, with
\begin{equation*}
  q(z_{i}) \propto \phi(z_{i} - m_{i}) \; \one\{z_{i}\in\mathcal{A}_{y_{i}}\},
\end{equation*}
where $m_{i} = \bx_{i}\T \bW \bmu$.
Thus, $q(z_{i})$ is the density of a Gaussian random variable $\Normal(m_{i},1)$ truncated to $\mathcal{A}_{y_{i}}$.
Letting $k_{i}=2y_{i}-1\in\{-1,1\}$, standard formulas for the moments of truncated Gaussian random variables give
\begin{align}
  \bar z_{i} &= \E_{q(z_{i})}[z_{i}] = m_{i} + k_{i}\,\lambda(k_{i} m_{i}), \label{eq:Ez}\\
  \E_{q(z_{i})}[z_{i}^{2}] &= 1 + m_{i}\,\bar z_{i}, \label{eq:Ez2}
\end{align}
where $\lambda(t)=\phi(t)/\Phi(t)$ denotes the inverse Mills ratio.
We only need $\bar{\bz}=(\bar z_{1},\ldots,\bar z_{n})\T$ in~\eqref{eq:betaVB}; the second moment in~\eqref{eq:Ez2} is useful for monitoring the evidence lower bound (ELBO), which is reported in the Supplementary Material.

The expressions \eqref{eq:Ez} and \eqref{eq:Ez2} follow from standard properties of truncated Gaussian distributions.
Indeed, let $T\sim \Normal(m,1)$ and condition on $T>0$.
Then, $\E[T] = m + \lambda(m)$ and $\Var(T) = 1 - m\,\lambda(m) - \lambda(m)^{2}$.
Hence $\E[T^{2}]=\Var(T)+\E[T]^{2}=1 - m\,\lambda(m) - \lambda(m)^{2} + (m+\lambda(m))^{2} = 1 + m\,\E[T]$.
In case we condition on $T\le 0$, we have $\E[T] = m - \lambda(-m)$ and $\Var(T) = 1 + m\,\lambda(-m) - \lambda(-m)^{2}$ instead.
Adapting the previous computations, we get $\E[T^{2}]=1 + m\,\E[T]$ also in this case.
Equations \eqref{eq:Ez} and \eqref{eq:Ez2} give a concise representation of these results.

\subsection{The optimal variational factor \texorpdfstring{$q(\gamma_{j})$}{q(gammaj)}}
\label{subsec:2.3}
First, notice that the full conditional of $\gamma_{j}$ takes the following form:
\begin{equation*}
\begin{split}
    p(\gamma_{j}\mid \bgamma_{-j},\bz,\bbeta,\by) &\propto p(\bbeta)p(\bgamma)p(\bz\mid\bbeta,\bgamma)p(\by\mid\bz)\\
    &\propto p(\bgamma)p(\bz\mid\bbeta,\bgamma)\\ 
    &\propto \rho^{\gamma_{j}}(1-\rho)^{1-\gamma_{j}}\exp\left\{ -\frac{1}{2} (\bz-\bX\bGamma\bbeta)\T (\bz-\bX\bGamma\bbeta) \right\},
\end{split}
\end{equation*}
and consequently
\begin{equation}
\label{eq:log_full_cond_gamma_j}
    \log p(\gamma_{j}\mid \bgamma_{-j},\bz,\bbeta,\by) = \gamma_{j}\text{logit}(\rho) -\frac{1}{2} \bbeta\T\bGamma\bX\T \bX\bGamma\bbeta + \bbeta\T\bGamma\bX\T\bz + \text{const}.
\end{equation}
Before isolating $\gamma_{j}$ in the full conditional and taking the variational expectation to compute $q(\gamma_{j})$, let us recall that if $V\sim \Bern(p)$ then $p(v)=p^v(1-p)^{1-v}\propto \left(\frac{p}{1-p}\right)^v\propto \exp\left\{v\cdot\text{logit}(p)\right\}$ and thus $\log p(v) = v\cdot \text{logit}(p) +\text{const}$.
Let us now isolate the term $\gamma_{j}$ in the second and third terms in \eqref{eq:log_full_cond_gamma_j}.
It holds $\bX\bGamma\bbeta = \gamma_{j} \beta_{j} \bX_{:j} + \sum_{j'\ne j} \gamma_{j'} \beta_{j'} \bX_{:j'}$, thus
\begin{equation*}
    \bbeta\T\bGamma\bX\T\bz = \gamma_{j} \beta_{j} \bX_{:j}\T\bz + \sum_{j'\ne j} \gamma_{j'} \beta_{j'} \bX_{:j'}\T\bz = \gamma_{j} \beta_{j} \bX_{:j}\T\bz + \text{const},
\end{equation*}
and
\begin{align*}
    \bbeta\T\bGamma\bX\T \bX\bGamma\bbeta &= \left(\sum_{l=1}^{p} \gamma_{l} \beta_{l} \bX_{:l}\right)\T \left(\sum_{k=1}^{p} \gamma_{k} \beta_{k} \bX_{:k}\right)\\
    &= \gamma_{j}\beta_{j}\bX_{:j}\T \sum_{k=1}^{p} \bX_{:k} \gamma_{k} \beta_{k} + \left(\sum_{l\ne j} \gamma_{l} \beta_{l} \bX_{:l}\T\right) \gamma_{j}\beta_{j}\bX_{:j} + \text{const}\\
    &= \gamma_{j} \beta_{j}^{2}\bX_{:j}\T\bX_{:j} + 2\gamma_{j}\beta_{j}\bX_{:j}\T \left(\sum_{l\ne j} \gamma_{l} \beta_{l} \bX_{:l}\right) + \text{const}\\
    &= \gamma_{j}\beta_{j}^{2}\bX_{:j}\T\bX_{:j} + 2\gamma_{j}\beta_{j}\bX_{:j}\T \bX_{-j}\bGamma_{-j,-j}\bbeta_{-j} + \text{const},
\end{align*}
where $\bX_{-j}$ corresponds to $\bX$ with the $j$-th column $\bX_{:j}$ removed, and $\bGamma_{-j,-j}$ is the matrix $\bGamma$ with the $j$-th row and the $j$-th column removed.
Putting it all together, we obtain
\begin{multline*}
    \log p(\gamma_{j}\mid \bgamma_{-j},\bz,\bbeta,\by)\\
    = \gamma_{j}\left[ \text{logit}(\rho) -\frac{1}{2} \beta_{j}^{2}\bX_{:j}\T\bX_{:j} + \beta_{j}\bX_{:j}\T(\bz - \bX_{-j}\bGamma_{-j,-j}\bbeta_{-j}) \right]
    + \text{const}.
\end{multline*}
Taking the expectation with respect to all the other variational distributions, we obtain that the optimal $q(\gamma_{j})$ must satisfy
\begin{equation*}
    \begin{split}
        \log q(\gamma_{j}) &= \E_{q(\bbeta) q(\bgamma_{-j}) q(\bz) }\left[ \log p(\gamma_{j}\mid \bgamma_{-j},\bz,\bbeta,\by) \right] + \text{const}\\
        &= \gamma_{j} \cdot \left[ \text{logit}(\rho) -\frac{1}{2} (\Sigma_{jj} + \mu_{j}^{2})G_{jj} + \mu_{j}\bX_{:j}\T\bar{\bz} - \sum_{k\ne j} (\Sigma_{jk} + \mu_{j}\mu_{k})w_{k} G_{jk} \right]\\
    &\qquad \qquad \qquad \qquad \qquad \qquad \qquad \qquad \qquad \qquad \qquad \qquad \qquad + \text{const},
    \end{split}
\end{equation*}
from which we obtain that the optimal variational factor $q(\gamma_{j})$ is the probability mass function of a $\Bern(w_{j})$ random variable, with $w_{j}=\expit(\eta_{j})$ and
\begin{equation}
\label{eq:eta_j}
    \eta_{j} = \logit(\rho) \;-\; \frac{1}{2}\, ( \Sigma_{jj}+\mu_{j}^{2} )\, G_{jj} \;+\; \mu_{j}\, \bX_{:j}\T \bar{\bz} \;-\; \sum_{k\neq j} ( \Sigma_{jk}+\mu_{j}\mu_{k} )\, w_{k}\, G_{jk}.
\end{equation}

\section{The algorithm}
\label{sec:3}
The optimal values of the variational parameters can be obtained via a coordinate ascent variational inference (CAVI) algorithm, where the parameters are sequentially updated until their maximum relative change or the change in the ELBO becomes negligible.
The routine is given in Algorithm~\ref{algo:1}.
\begin{algorithm}[H]
\caption{MFVB for sparse probit regression}
\begin{algorithmic}[1]
\State \textbf{Input:} $\bX\in\R^{n\times p},\,\by\in\{0,1\}^{n}$, $\nu^{2}$, $\rho$.
\State \textbf{Precompute} $\bG \gets \bX\T \bX$.
For $i=1,\dots,n$: $k_{i}\gets 2y_{i}-1$.
\State \textbf{Initialize} $\bw$ and $\bmu$ (e.g., set $w_{j}\gets\rho$, $\mu_{j}\gets 0$ for $j=1,\ldots,p$).
\State \textbf{Set} $\bW \gets \diag(\bw)$;\ \ $\bvec m \gets \bX \bW \bmu$;\ \ for $i=1,\dots,n$: $\bar z_{i} \gets m_{i} + k_{i}\,\lambda(k_{i} m_{i})$.
\Repeat
  \State $\bOmega \gets \bW(\bI_{p}-\bW)+\bw\bw\T$
  \State $\bSigma \gets (\nu^{-2}\bI_{p} + \bG\Had \bOmega)^{-1}$;\quad $\bmu \gets \bSigma \bW \bX\T \bar{\bz}$
  \State $\bvec m \gets \bX \bW \bmu$;\quad for $i=1,\dots,n$: $\bar z_{i} \gets m_{i} + k_{i}\,\lambda(k_{i} m_{i})$
  \State For $j=1,\dots,p$: compute $\eta_{j}$ by~\eqref{eq:eta_j} and set $w_{j}\gets \expit(\eta_{j})$
  \State $\bW \gets \diag(\bw)$
\Until{the relative change in the ELBO or in $(\bmu,\bw)$ is below the tolerance}
\State \textbf{Return:} $(\bmu,\bSigma)$, inclusion probabilities $\bw$, and ELBO.
\end{algorithmic}
\label{algo:1}
\end{algorithm}
We now discuss some computational and practical considerations about the algorithm.
\subsection[Efficient approximations for large p]{Efficient approximations for large $p$}
Algorithm~\ref{algo:1} updates the Gaussian factor $q(\bbeta)$ with precision matrix $\bSigma^{-1}=\nu^{-2}\bI_{p}+\bG\Had\bOmega$, where $\bG$ and $\bOmega$ are defined above.
A direct computation of $\bSigma$ costs $\mathcal{O}(p^{3})$ and can be prohibitive for large $p$.
In practice, when many PIPs degenerate close to $0$ or when $p>n$, one can avoid this cost using the following simplifications (detailed in the Supplementary Material).
First, when many inclusion probabilities $w_{j}$ are close to zero, only a subset of predictors contributes to the linear predictor; one can therefore restrict updates to an active set $S_{\bw}=\{j:w_{j}>\varepsilon\}$ of size $k$, treating $w_{j} \le \varepsilon$ (e.g., $\varepsilon=10^{-10}$) as zero and keeping the corresponding coefficients at their prior values.
The update of $q(\bbeta)$ is then performed on the reduced $k$-dimensional system: if $k\le n$ we work with the $k\times k$ restricted precision matrix, while if $k>n$ we apply Woodbury's identity so that the dominant linear algebra involves an $n\times n$ matrix.
Finally, in very high dimensions, we can approximate the update of $q(\gamma_{j})$ by using an approximate version of $\eta_{j}$ that avoids off-diagonal terms of $\bSigma$, yielding additional computational savings.
Note that in this way, one obtains a variational algorithm between our default (with full-covariance $q(\bbeta)$) and a fully factorized VB that also assumes a more restrictive independence structure for the multivariate factor $q(\bbeta)$.

\subsection{Initialization and stopping rule}
Iterations are run until convergence, defined as the maximum relative change between two successive iterations in the variational coefficient means and the inclusion probabilities, measured in Euclidean norm, falling below a prescribed tolerance.
In our experiments, we use a tolerance of $10^{-4}$.
Note that in all experiments on synthetic and real data, we initialize the variational parameters at $\mu_{j}=0$ and $w_{j}=\rho$ for all $j=1,\ldots,p$ (Algorithm~\ref{algo:1}, line~3).
Setting $\mu_{j}=0$ is the neutral starting point that does not induce an initial directional bias in the latent mean vector $\bvec m = \bX\bW\bmu$ and hence in the sign of $\bar{\bz}$, consistent with the prior specification.
Initializing $w_{j}$ at $\rho$ is also natural, as it matches the prior inclusion expectation.
In CAVI optimization, different initializations can, in principle, lead to different local optima.
As a practical check, one can run the algorithm from multiple starting values for $\bw$ and retain the solution with the largest final ELBO.
In our numerical studies, the default initialization above provided stable solutions and we did not observe meaningful changes in the main inferential and predictive summaries for small perturbations of such initializations.
Moreover, when the posterior places similar mass on multiple, nearly equivalent variable selection configurations (for instance, under weak signals or highly correlated predictors), a variational approximation may concentrate on one of these nearly equivalent configurations and yield overconfident PIPs.
In such cases, different initializations may select different but practically equivalent subsets, typically with comparable ELBO values and predictive performance, reflecting practical non-identifiability rather than substantive instability in the fitted model.
Therefore, while this behavior can affect Bayesian uncertainty quantification over model space, it is less problematic when the goal is prediction or point estimation of a selected sparse model.

\subsection{Hyperparameter settings}
To run the algorithm, one needs to set the two hyperparameters $\nu^{2}$ (prior variance for $\bbeta$) and $\rho$ (prior inclusion probability of the variables).
In both the simulation studies and the applications below, these hyperparameters are set according to the following data-driven procedure.
First, for a given $\rho$ value, we set $\nu^{2}=\nu_{0}^{2}/(\rho\cdot p)$, in order to control the prior variance of the linear predictor to be close to $\nu_{0}^{2}$, under the assumption that the predictors have been standardized (as it is always advisable in practice) or generated with unit variance.
The role of $\nu_{0}^{2}$ is to control the prior dispersion of the linear predictor and thus also of the implied prior predictive probabilities.
Under standardized predictors and the prior inclusion rate $\rho$, the rescaling $\nu^{2}=\nu_{0}^{2}/(\rho p)$ yields $\Var(\bx_{i}\T \bGamma \bbeta)\approx \nu_{0}^{2}$ a priori.
In the following, we set $\nu_{0}^{2}=25$, corresponding to a standard deviation of about $5$ on the latent probit scale, which provides a weakly informative prior that allows substantial effect sizes while avoiding an overly concentrated prior around zero, following common suggestions in Bayesian binary regression models \citep{gelman2008weakly, fasano2022scalable}.
We also report a sensitivity analysis over a range of $\nu_{0}^{2}$ values in the Supplementary Material.

Then, after potentially dividing the data into a training and test set, the optimal value of $\rho$ is chosen via a $K$-fold cross-validation on the training set only, where the allocation of observations to folds is stratified by output value $y_{i}$ to have approximately the same fraction of $y_{i}=1$ in each fold, close to the fraction in the whole training dataset.
For each value of $\rho$ in a grid of possible values, typically ranging from $0.05$ to $0.5$ with a step size of $0.05$, we compute the average cross-validation deviance
\begin{equation}
\label{eq:devCV}
    \text{dev}_{CV} = \frac{1}{K}\sum_{k=1}^K \text{dev}_{k},
\end{equation}
where $\text{dev}_{k}= -2\sum_{i\in I_{k}}[y_{i} \log \hat p_{i} + (1-y_{i}) \log (1-\hat p_{i})]$, $I_{k}$ is the set of indices of the observations assigned to fold $k$, and $\hat p_{i} = \hat\Pr[y_{i}=1\mid \by_{-\text{FOLD}(i)}]$ is the estimated probability of success for the $i$-th training observation, when the model is fitted on the set of all the training observations, excluding the ones in the fold to which the $i$-th observation belongs, whose index set is denoted $\text{FOLD}(i)=I_{k} \text{ if }i\in I_{k}$.
Calling $\bW^{(-\text{FOLD}(i))}$ and $\bmu^{(-\text{FOLD}(i))}$ the variational mean estimates for $\bGamma$ and $\bbeta$ obtained from the MFVB algorithm fitted on $\by_{-\text{FOLD}(i)}$, we compute $\hat p_{i}$ using the plug-in estimate $\hat p_{i}=\Phi(\bx_{i}\T \bW^{(-\text{FOLD}(i))} \bmu^{(-\text{FOLD}(i))})$.

\section{Simulation studies}
\label{sec:4}
\begin{figure}[b!]
    \centering
    \includegraphics[width=\linewidth]{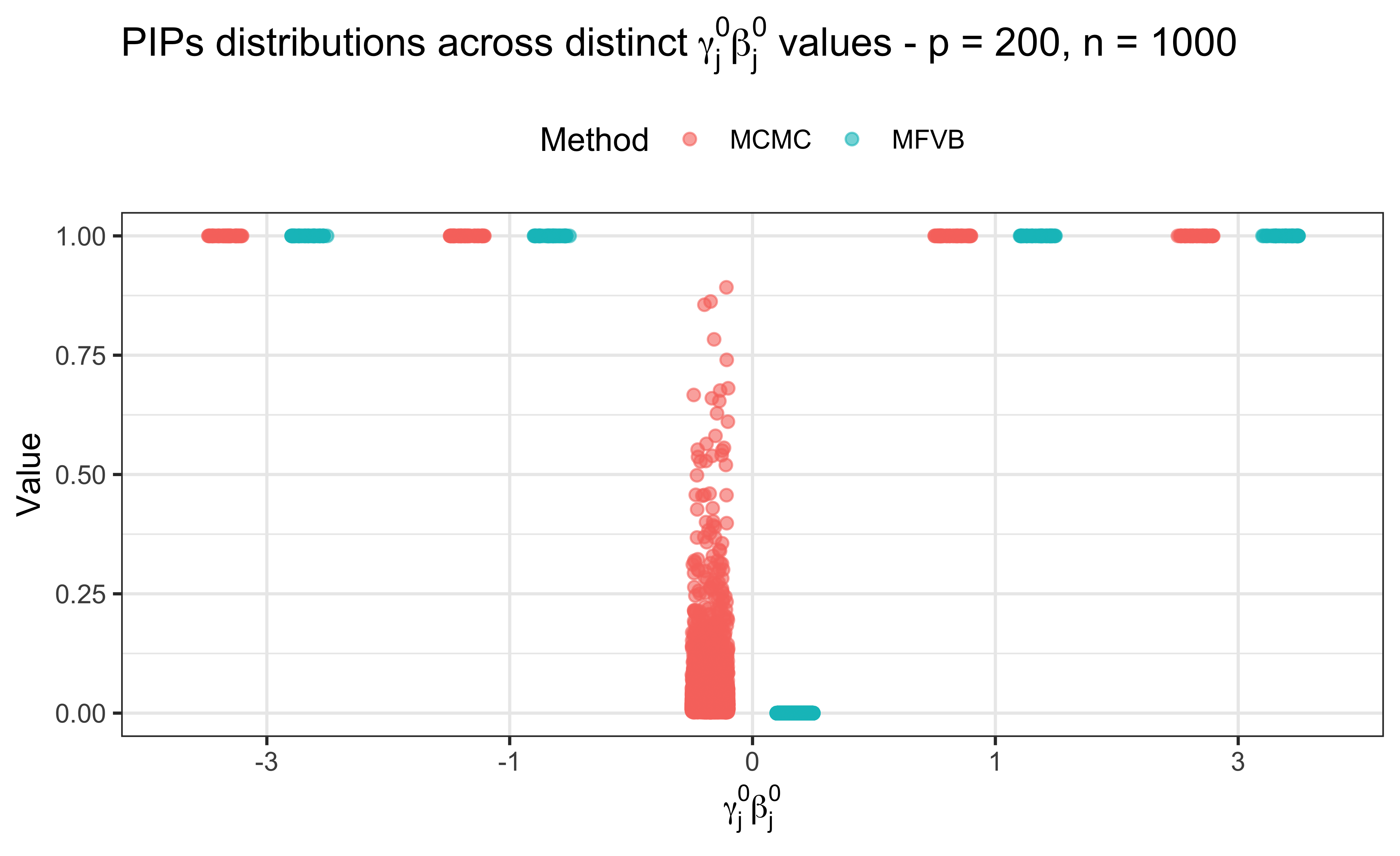}
    \caption{For $p=200$, $n=1000$, posterior inclusion probabilities (PIPs) as a function of the true parameter values $\gamma_{j}^{0}\beta_{j}^{0}$ estimated by MCMC and MFVB across the $50$ simulated datasets.
    For graphical purposes, the distance between $0$ and $1$ is not to scale and we spread the true $\gamma_{j}^{0}\beta_{j}^{0} \in \{\pm 3, \pm 1, 0\}$ in a neighborhood of their actual values.}
    \label{fig:fig1}
\end{figure}

\begin{figure}[t]
    \centering
    \includegraphics[width=\linewidth]{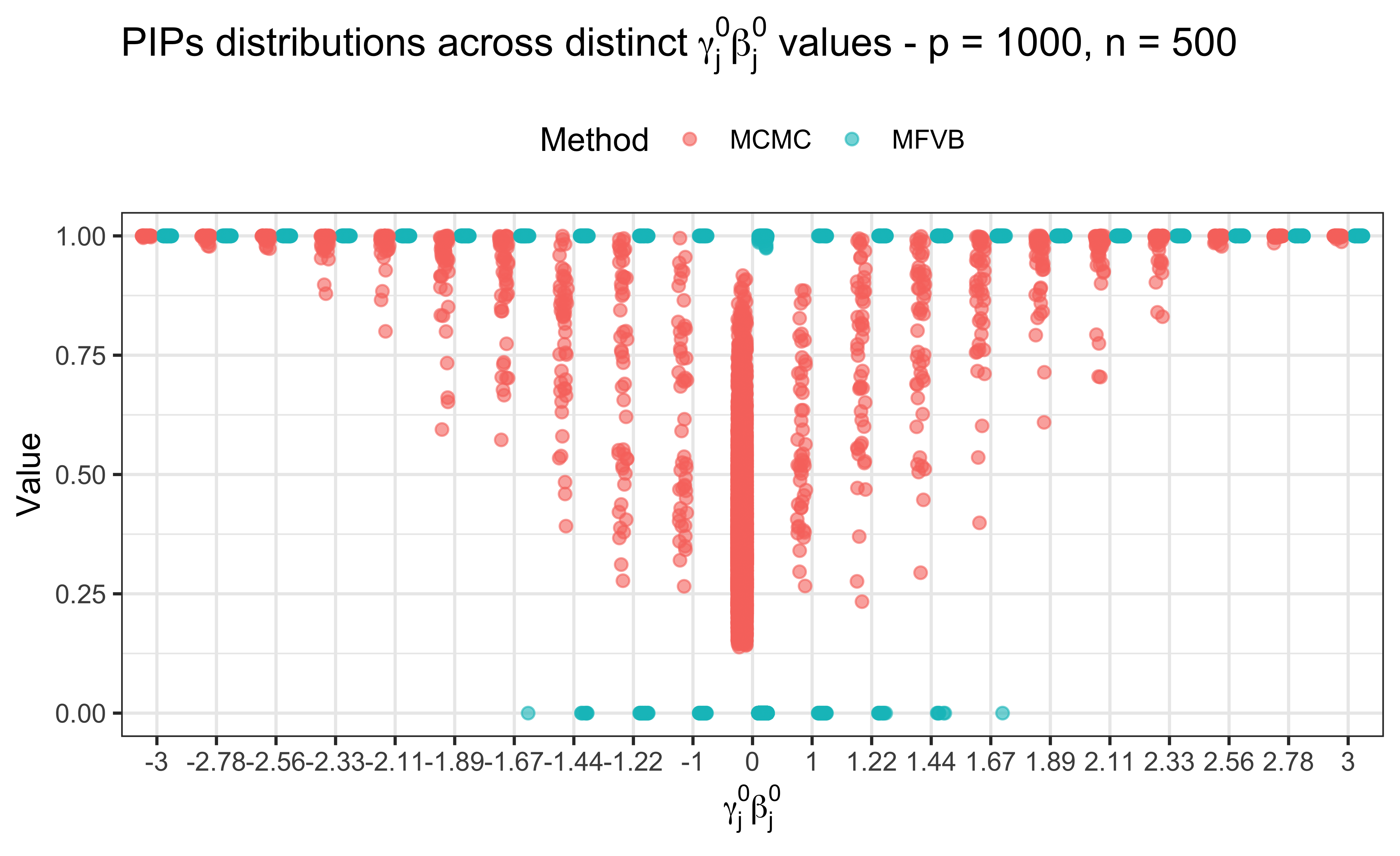}
    \caption{For $p=1000,\ n=500$, posterior inclusion probabilities (PIPs) as a function of the true parameter values $\gamma_{j}^{0}\beta_{j}^{0}$ estimated by MCMC and MFVB across the $50$ simulated datasets.
    For graphical purposes, the distance between $0$ and $1$ is not to scale and we spread the true regression parameters $\gamma_{j}^{0}\beta_{j}^{0}$ in a neighborhood of their actual values.}
    \label{fig:fig2}
\end{figure}

We investigate the performance of the proposed algorithm in selecting the appropriate variables and compare its performance with MCMC sampling methods, where $10\,000$ posterior samples (after a burn-in of $1000$) are obtained from a Gibbs sampler, whose details are reported in the Supplementary Material and which leverages the expressions of the full conditionals.
We consider two different scenarios: one where $p=200$ and $n=1000$ and a more challenging one where $p=1000$ and $n=500$.
We consider $50$ replicated datasets in each scenario.
In both cases, in each replication, the true number of non-zero coefficients is $0.02\cdot p$, with half of them being equally-spaced points between $-3$ and $-1$ and the other half being equally-spaced points between $1$ and $3$.
The covariate values are i.i.d.\ standard normal and the resulting output values are generated according to model \eqref{eq:probit_model}.
For each dataset, we compute the posterior approximation of the parameters with both MFVB and MCMC and compare the results.
We fix $\nu^{2}$ and $\rho$ as explained in Section~\ref{sec:3}, with $K=5$ folds in the cross-validation procedure and $\nu_{0}^{2}=25$.
We then compare the methods in terms of running times, true positive and negative rates, as well as predictive accuracy on a test set of size $500$, different from the training set used to tune $\rho$.
For posterior comparison, we consider a variable $j$ to be included if the estimated posterior inclusion probability $\hat\Pr[\gamma_{j}=1\mid \by]$ exceeds $0.5$; otherwise, the variable is excluded.
We compute the true positive rate as the number of correctly included variables over the total number of active variables, while the true negative rate is the number of correctly excluded variables over the total number of inactive variables.
Predictive probabilities for a new observation $y_{\new}$ with covariate vector $\bx_{\new}$ for MFVB are computed using the plug-in estimate $\hat\Pr[y_{\new}=1\mid \by]=\Phi(\bx_{\new}\T \bW\bmu)$, where, consistent with the notation above, $\bW$ is the diagonal matrix with posterior expectations of $\bgamma$ and $\bmu$ represents the posterior expectations of $\bbeta$ obtained with MFVB.
For MCMC, predictive probabilities are computed by averaging over posterior draws.
All simulations were executed on a 12-core Mac mini featuring an Apple M4 Pro processor.
\begin{table}[ht]
\centering
\footnotesize
\caption{Mean results by scenario and method.
Standard deviations in parentheses.
\label{tab:results}}
\begin{tabular}{lcccc}
    \toprule
    & \multicolumn{2}{c}{$p=200$, $n=1000$} & \multicolumn{2}{c}{$p=1000$, $n=500$} \\
    \cmidrule(lr){2-3} \cmidrule(lr){4-5}
    \textbf{Metric} & MFVB & MCMC & MFVB & MCMC\\
    \midrule
    True positive rate (\%) & 100 (0) & 100 (0) & 90.30 (7.17) & 94.40 (4.70) \\
    True negative rate (\%) & 100 (0) & 99.77 (0.35) & 99.78 (0.23) & 94.95 (3.60) \\
    Deviance & 161.31 (15.44) & 163.51 (15.92) & 219.15 (25.59) & 441.51 (29.47) \\
    Running time (secs) & 0.93 (0.24) & 48.54 (2.53) & 27.34 (2.58) & 60958.84 (24854.62)\\
    Effective sample size $\beta_{j}$ & --- & 2040.53 (542.30) & --- & 1303.17 (187.32)\\
    Effective sample size $\gamma_{j}$ & --- & 3394.44 (567.99) & --- & 3399.19 (703.90)\\
\bottomrule
\end{tabular}
\end{table}
Table~\ref{tab:results} also reports the average effective sample size for $\beta_{j}$ and $\gamma_{j}$ as a standard mixing diagnostic for the Gibbs sampler.
As shown in Table~\ref{tab:results}, in both $p < n$ and $p > n$ settings, the MFVB is more than one order of magnitude faster in the first scenario, and over three orders in the second one.
Moreover, in the simpler $p\ll n$ setting, the two methods provide practically equal results in terms of variable selection and out-of-sample deviance, with an almost perfect variable selection.
In the challenging $p \gg n$ setting, both methods perform well in terms of true positive and true negative rates, although some small differences are present between the two methods.
These differences can be better understood by looking at the estimated posterior inclusion probabilities (PIPs hereafter) plotted as a function of the data-generating regression parameters $\gamma_{j}^{0}\beta_{j}^{0}$ across repeated simulations, for both the first (Figure~\ref{fig:fig1}) and second (Figure~\ref{fig:fig2}) scenario.
In general, we observe that MFVB PIPs tend to degenerate toward the boundaries $0$ or $1$, in line with theoretical results in \citet{ormerod2017variational} for linear models, where it is shown that, under proper initialization, they converge to the true data-generating indicators $\gamma_{j}^{0}\in\{0,1\}$.
On the other hand, MCMC tends to be less sparse, possibly including more variables than needed: especially for the $p\gg n$ scenario, this may result in a higher out-of-sample deviance.

Considering now the two scenarios more in detail, in the first ($p\ll n$) one, the MFVB estimates also match the PIP estimates obtained with MCMC for values of $\gamma_{j}^{0}\beta_{j}^{0}$ different from zero.
See Figure~\ref{fig:fig1}.
Indeed, in such a scenario, the high number of observations compared to the dimension of the problem allows both methods to learn well which are the truly active variables.
As for the inactive variables, i.e., the ones where the data-generating parameter $\gamma_{j}^{0}$ equals zero, the MCMC tends to be more conservative, with multiple estimated PIPs away from the boundaries, while MFVB still shows an almost deterministic classification, as anticipated.
This peculiar property seems to be beneficial for MFVB in high-dimensional settings, especially when $p\gg n$, as in the case $p=1000$, $n=500$ considered in the second scenario, with results reported in Figure~\ref{fig:fig2}.
In such a scenario, the PIPs estimated by the two methods coincide for active variables whose true coefficient $\beta_{j}^{0}$ is large in absolute value.
For active variables whose true coefficient $\beta_{j}^{0}$ has values not sufficiently far away from zero, the MCMC estimates are more dispersed, although there is a general agreement among the two methods on which variables should be included.
Finally, for inactive variables, the MCMC seems more conservative, giving higher PIPs.
In practice, we observe that MFVB favors more parsimonious representations, while MCMC, being more conservative, tends to be more prone to overfitting the data.
Thus, the degeneracy of the PIPs observed for MFVB might seem a favorable property, possibly useful in settings where $p$ is (much) larger than $n$.

While the near-degenerate PIPs can yield a clear ranking, a parsimonious selection and efficient computation, it is also consistent with a known limitation of mean-field variational approximations, namely an underestimation of posterior uncertainty.
In particular, the factorization assumed for the inclusion indicators can produce overly concentrated variational posteriors, so that PIPs may be closer to $0$ or $1$ than under MCMC.
This mainly affects Bayesian uncertainty quantification (for example, credible intervals or posterior dispersion of inclusion indicators) and may lead to overconfident inclusion or exclusion decisions when signals are weak or predictors are highly correlated.
For prediction or point estimation of the selected model, we did not observe adverse effects in our experiments, but we caution against interpreting MFVB PIPs as fully calibrated posterior probabilities.
Note that the MFVB independence assumption mainly concerns the inclusion indicators, whereas the variational distribution of the regression coefficients is not factorized.
Since $q(\bbeta)$ retains a full covariance structure, the approximation can represent posterior dependence among coefficients associated with correlated predictors.
\FloatBarrier

\section{Real-data applications}\label{sec:5}
We evaluate the proposed method on two real-data applications: the \emph{LSVT Voice Rehabilitation} dataset, first analyzed by \citet{tsanas2013objective}, and the \emph{Alzheimer} dataset, studied in \citet{craig2011multiplexed} and further considered in \citet{fasano2022scalable}.
The former dataset has $n=126$ observations and, after cleaning, $p=309$ predictors, with the outcome representing the acceptability of the vocal performance of patients with Parkinson's disease.
The latter addresses a large-$p$ scenario: it consists of $n=333$ observations on the early onset of Alzheimer's disease, with associated $p=9036$ predictors (obtained by adding pairwise interaction terms in the original dataset), showing the practical usefulness of the proposed approach in settings where MCMC is not computationally feasible.
In both applications, we apply the MFVB procedure described in Section~\ref{sec2}, with $\rho$ and $\nu^{2}$ set as explained in Section~\ref{sec:3} and $\nu_{0}^{2}=25$, using the whole dataset as a training set.

\subsection{LSVT Voice Rehabilitation application}
\label{subsec:5.1}
We first evaluate the proposed MFVB approach on the \emph{LSVT Voice Rehabilitation} dataset, which concerns the classification of vocal performance of patients with Parkinson's disease (PD) as ``acceptable'' ($y_{i}=1$) or ``unacceptable'' ($y_{i}=0$).
The dataset comprises $n=126$ phonations from $14$ patients and is publicly available at the UCI repository.\footnote{\url{https://archive.ics.uci.edu/dataset/282/lsvt+voice+rehabilitation}}
First analyzed by \citet{tsanas2013objective}, the dataset contains $310$ features, of which two (\texttt{Data\_length} and \texttt{Ea2}) show very limited variability and are thus removed from the analysis.
The remaining covariates are standardized to a zero mean and unit variance.
An intercept is also added, yielding $p=309$ predictors.
The original aim in \citet{tsanas2013objective} was to replicate experts' ratings with automated predictors and to identify influential features: our analysis targets the same goals with a fully Bayesian, sparse, and interpretable model.

\begin{table}[t]
\centering
\caption{For both MFVB and MCMC, included features ranked by posterior inclusion probability (PIP) for the LSVT Voice Rehabilitation application: MFVB vs.\ MCMC.
Bold entries represent the features for which the MFVB estimated PIP $w_{j}$ is greater than~$0.5$.}
\label{table:2}
\centering
\fontsize{8}{11}\selectfont
\begin{tabular}[t]{lrrlrr}
\toprule
Feature (MFVB) & PIP $(w_{j})$ & $w_{j} \mu_{j}$ & Feature (MCMC) & PIP & $\mathbb{E}[\gamma_{j}\beta_{j}]$\\
\midrule

\textbf{Intercept} & 1.000 & -0.743 & \textbf{Intercept} & 1.000 & -1.280\\
\textbf{IMF}\verb|->|\textbf{NSR\_SEO} & 1.000 & 0.693 & \textbf{MFCC\_0th coef} & 0.742 & -0.429\\
\textbf{Shimmer}\verb|->|\textbf{Ampl\_abs0th\_perturb} & 1.000 & -0.599 & \textbf{MFCC\_1st coef} & 0.681 & 0.351\\
\textbf{MFCC\_0th coef} & 1.000 & -0.576 & \textbf{IMF}\verb|->|\textbf{NSR\_SEO} & 0.679 & 0.358\\
\textbf{MFCC\_1st coef} & 1.000 & 0.501 & \textbf{MFCC\_7th coef} & 0.670 & 0.294\\
\textbf{HNR}\verb|->|\textbf{HNR\_dB\_Praat\_std} & 1.000 & -0.415 & 12th delta & 0.669 & -0.284\\
\textbf{MFCC\_12th coef} & 0.997 & 0.311 & MFCC\_4th coef & 0.658 & 0.306\\
\textbf{MFCC\_7th coef} & 0.993 & 0.285 & MFCC\_8th coef & 0.645 & 0.273\\
 & & & DFA & 0.641 & 0.314\\
 & & & IMF\verb|->|NSR\_entropy & 0.632 & 0.298\\
 & & & Log energy & 0.605 & -0.270\\
 & & & MFCC\_9th coef & 0.594 & -0.233\\
 & & & Jitter\verb|->|pitch\_TKEO\_prc75 & 0.578 & 0.227\\
 & & & entropy\_log\_5\_coef & 0.569 & -0.228\\
 & & & \textbf{HNR}\verb|->|\textbf{HNR\_dB\_Praat\_std} & 0.562 & -0.209\\
 & & & entropy\_log\_2\_coef & 0.549 & -0.204\\
 & & & entropy\_log\_4\_coef & 0.546 & -0.206\\
 & & & IMF\verb|->|SNR\_TKEO & 0.545 & 0.217\\
 & & & GNE\verb|->|std & 0.520 & -0.181\\
 & & & entropy\_log\_7\_coef & 0.519 & 0.171\\
 & & & \textbf{Shimmer}\verb|->|\textbf{Ampl\_abs0th\_perturb} & 0.512 & -0.170\\
 & & & VFER\verb|->|NSR\_SEO & 0.508 & 0.163\\
 & & & entropy\_log3\_4\_coef & 0.508 & -0.160\\
 & & & IMF\verb|->|NSR\_TKEO & 0.505 & 0.155\\
 & & & Shimmer\verb|->|Ampl\_AM & 0.504 & -0.160\\
\bottomrule
\end{tabular}
\end{table}

We compare the proposed MFVB approach, with hyperparameters set as above, to the Gibbs sampling baseline (MCMC) under the same prior specification.
Table~\ref{table:2} reports the variables included by MFVB and MCMC, ranked by posterior inclusion probability (PIP).
As shown in the table, MFVB includes (i.e., returns $w_{j}>0.5$ for the corresponding variable) the intercept and seven additional variables, highlighted with boldface, while it assigns negligible inclusion probabilities to the remainder.
Of these eight selected variables, five are among the five most influential variables according to MCMC and two additional variables appear in the list of variables selected by the MCMC.
Thus, seven of the eight variables singled out by MFVB are corroborated by the sampler, while MFVB remains more parsimonious.
Indeed, MCMC tends to include more variables, assigning $\text{PIP}>0.5$ to $25$ variables, a pattern consistent with the simulations reported in Section~\ref{sec:4}.

\FloatBarrier
Considering the out-of-sample predictive accuracy, we obtain an estimate via $5$-fold cross-validation (with different folds from the ones used to optimize $\rho$), where, for each of the $5$ folds, we compute the predictive probabilities for observations in that fold by training the model on the remaining four, obtaining a fold average deviance as in equation \eqref{eq:devCV}.
Taking the average over folds, this resulted in comparable mean deviance values: $18.89$ for MFVB versus $19.29$ for MCMC.
So, the configuration identified with MFVB is expected to have comparable out-of-sample performance, but it is more parsimonious and could potentially better highlight the most important variables.
Moreover, the MFVB took $0.16$ seconds to run, while the MCMC sampler required around $544$ seconds, three orders of magnitude higher.
In terms of classification accuracy, predicting $\hat y_{\new}=1$ whenever the posterior predictive probability for a new observation with covariate vector $\bx_{\new}$ exceeds $0.5$, we obtain $86.5\%$ for MFVB and $82.5\%$ for MCMC, which is comparable to the levels reported for more sophisticated machine learning techniques in \citet{tsanas2013objective}, with the advantage of a fully interpretable model.
Note that the deviance (reported above) is based on the full predictive probabilities, whereas the misclassification rate depends on a threshold (here $0.5$).
Overall, the proposed approach provides a coherent, unified framework to jointly perform variable selection and prediction, combining interpretable results and a predictive accuracy comparable to more advanced black-box routines.
\FloatBarrier

\subsection{Alzheimer's disease application}
\label{subsec:5.2}

The second real-data application focuses on a large-$p$ setting where MCMC approaches are not feasible.
In particular, the application is built starting from the \texttt{AlzheimerDisease} dataset available in the \texttt{R} package \texttt{Applied\-PredictiveModeling} \citep{kuhn2018applied}.
The dataset, first analyzed in \citet{craig2011multiplexed}, concerns the presence or absence of Alzheimer's disease, encoded here as $y_{i}=1$ or $y_{i}=0$, respectively, for $n=333$ patients, as a function of demographic data and laboratory results.
Following \citet{fasano2022scalable}, the original dataset is modified by adding all pairwise interactions among predictors, together with the intercept, resulting in $p=9036$ predictors.
For such a high value of $p$, MCMC is not feasible, and the proposed MFVB could also face computational issues if Algorithm~\ref{algo:1} were implemented naively.
We thus exploit some computational simplifications by setting the covariance terms $\Sigma_{jk}=0$ for $j\ne k$ in the evaluation of $\eta_{j}$ in equation \eqref{eq:eta_j}, thereby enforcing independence across the $\beta_{j}$'s only in the $\gamma$-update.
Details are reported in Section~\ref{app:impl} of the Supplementary Material.
These simplifications reduce the running time of MFVB to $27$ seconds.
The prior hyperparameters $\nu_{0}^{2}$ and $\rho$ are set as in the LSVT Voice Rehabilitation analysis.
\begin{table}[b]
\centering
\caption{Features included by MFVB, ranked by posterior inclusion probability, for the Alzheimer's disease application}
\label{table:3}
\centering
\fontsize{9}{11}\selectfont
\begin{tabular}[t]{lrr}
\toprule
Feature (MFVB) & $w_{j}$ & $w_{j} \mu_{j}$\\
\midrule
Intercept & 1.000 & -0.547\\
tau & 1.000 & 0.600\\
Ab\_42 & 1.000 & -0.424\\
VEGF & 1.000 & -0.400\\
GRO\_alpha & 0.999 & 0.340\\
Pancreatic\_polypeptide & 0.997 & 0.325\\
\bottomrule
\end{tabular}
\end{table}
As shown in Table~\ref{table:3}, five predictors, plus the intercept, are selected, according to the usual inclusion criterion based on PIPs exceeding $0.5$, with non-selected predictors having estimated PIPs below $0.1$.
The five predictors are the cerebrospinal fluid (CSF) tau (\texttt{tau}), the CSF amyloid-$\beta 42$ (\texttt{Ab\_42}), the Vascular Endothelial Growth Factor (\texttt{VEGF}), the Growth-Regulated alpha protein (\texttt{GRO\_alpha}), and the pancreatic polypeptide (\texttt{Pancreatic\_polypeptide}).
These results further corroborate the scientific literature on the topic, in particular the original study by \citet{craig2011multiplexed}, where the five predictors are included as important variables in all the machine-learning approaches considered, with the only exception of \texttt{GRO\_alpha} for the Boosted Tree algorithm.
Also, the signs of the estimated effects $\mu_{j}$ are consistent with the findings of the original paper.
Interestingly, the main discrepancy with the variables included by the multiple methods considered in \citet{craig2011multiplexed} concerns the Log Matrix Metalloproteinase-10 (\texttt{MMP-10}) predictor, which is not assigned a relevant PIP by the proposed MFVB, while it was included by the methods explored in \citet{craig2011multiplexed}.
A plausible explanation for this discrepancy lies in the strong correlation, namely $0.458$, of \texttt{MMP-10} with \texttt{tau} documented in \citet{craig2011multiplexed}.
In the sparse Bayesian framework considered here, once \texttt{tau} is assigned a high posterior inclusion probability, the additional contribution of \texttt{MMP-10} may be largely absorbed, rendering its marginal inclusion probability negligible.
This behavior is in line with the tendency of the spike-and-slab prior to favor parsimonious representations by discarding variables whose predictive signal is already largely captured by included ones.
Some further support for this interpretation may be found in the ROC analyses of Table 4 in \citet{craig2011multiplexed}, where, in a related setting, adding \texttt{MMP-10} to the log \texttt{tau}/\texttt{Ab\_42} ratio yields only a modest improvement in area under the curve (AUC), from 0.8443 to 0.8518, while other predictors selected by the proposed method, such as \texttt{VEGF}, provide a comparatively larger gain (AUC 0.8766).
Further results on the mean out-of-sample deviance and sensitivity analyses for the choice of $\nu_{0}^{2}$ are reported in the Supplementary Material.

\FloatBarrier

\section{Conclusion}
\label{sec:6}
In this paper, we presented a mean-field variational Bayes (MFVB) method for sparse binary classification problems with a probit link.
Sparsity is induced thanks to the use of a spike-and-slab prior on the coefficients; see also \citet{ormerod2017variational}.
As shown in simulated and real data applications, the proposed CAVI algorithm can be easily implemented thanks to closed-form updates and is orders of magnitude faster than MCMC procedures, leading to massive computational gains.
When the number of observations $n$ is sufficiently large relative to $p$, both MFVB and MCMC select the same variables and lead to comparable solutions both in terms of parameter estimates and predictive probabilities.
In scenarios where $p>n$, MFVB tends to favor more parsimonious solutions, with fewer variables selected, while MCMC tends to be more conservative, including more variables, although with smaller associated posterior inclusion probabilities (PIPs).
Further research directions could consider more structured approximations, as done in \citet{fasano2022scalable} for Bayesian probit models with a Gaussian prior, or more complex models, including, e.g., mixed-effects, where accurate and scalable approximations are of utmost importance \citep{zhou2024scalable}.
The code and the data to reproduce the results are available in the GitHub repository \texttt{augustofasano/MFVB\_Probit\_Variable\_Selection}\footnote{\url{https://github.com/augustofasano/MFVB\_Probit\_Variable\_Selection}}.

\newpage

\printbibliography

\end{refsection}

\begin{refsection}

\newpage

\clearpage\pagebreak\newpage
\pagestyle{fancy}
\fancyhf{}
\rhead{\bfseries\thepage}
\lhead{\bfseries SUPPLEMENTARY MATERIALS}

\baselineskip=27pt
\begin{center}
{\LARGE{Supplementary Materials for\\} 
\bf   Mean-field Variational Bayes for Sparse Probit Regression
}
\end{center}

\baselineskip=14pt
\vskip 2mm

\begin{center}
		Augusto Fasano and Giovanni Rebaudo
        \vskip 3mm
        ESOMAS Department, University of Torino and Collegio Carlo Alberto
\end{center}

\setcounter{equation}{0}
\setcounter{page}{1}
\setcounter{table}{1}
\setcounter{figure}{0}
\setcounter{section}{0}
\numberwithin{table}{section}
\renewcommand{\theequation}{S.\arabic{equation}}
\renewcommand{\thesubsection}{S.\arabic{section}.\arabic{subsection}}
\renewcommand{\thesection}{S.\arabic{section}}
\renewcommand{\theThm}{S.\arabic{Thm}}
\renewcommand{\theCor}{S.\arabic{Cor}}
\renewcommand{\theProp}{S.\arabic{Prop}}
\renewcommand{\theLem}{S.\arabic{Lem}}
\renewcommand{\thepage}{S.\arabic{page}}
\renewcommand{\thetable}{S.\arabic{table}}
\renewcommand{\thefigure}{S.\arabic{figure}}

\baselineskip=14pt 
\vskip 10mm

\section{Closed-form ELBO}\label{app:elbo}
Write the evidence lower bound (ELBO) as $\mathcal{L}=\E_{q}[\log p(\by,\bz,\bbeta,\bgamma)] - \E_{q}[\log q(\bz,\bbeta,\bgamma)]$.
Notice that the joint density of model \eqref{eq:probit_model} takes the form
\begin{equation*}
  p(\by,\bz,\bbeta,\bgamma) = p(\by\mid\bz) \; p(\bz\mid\bbeta,\bgamma) \; p(\bbeta) \; p(\bgamma),
\end{equation*}
where $p(\by\given\bz)=\prod_{i} \one\{y_{i}=\one(z_{i}>0)\}$ simply restricts $z_{i}$ to the appropriate half-line.
Because $q(\bz)$ is supported on $\mathcal{A}_{\by}$, $\E_{q}[\log p(\by\mid \bz)]=0$.
Recall that $\mathcal{A}_{y_{i}}=(0,\infty)$ if $y_{i}=1$ and $(-\infty,0]$ if $y_{i}=0$, and $\mathcal A_{\by}=\mathcal A_{y_{1}}\times\cdots\times\mathcal A_{y_{n}}$.
Define $S_{zz}=\sum_{i=1}^{n} \E_{q(z_{i})}[z_{i}^{2}]= \sum_{i=1}^{n} (1+m_{i}\bar z_{i})$ and $\bC_{\bbeta} = \bSigma+\bmu\bmu\T$.
Here $m_{i}=\bx_{i}^{\top} \bW \bmu$, $\bar z_{i}=\E_{q(z_{i})}[z_{i}]$, $k_{i}=2 y_{i}-1$ and $\lambda(t)=\phi(t)/\Phi(t)$ denotes the inverse Mills ratio, where $\phi$ and $\Phi$ denote the standard normal density and cdf (as in the main manuscript).
Then,
\begin{align}
    \E_{q}[\log p(\bz\mid \bbeta,\bgamma)] &= -\tfrac{n}{2}\log(2\pi) \;-\; \tfrac{1}{2}\Big\{ S_{zz} - 2\,\bmu\T \bW \bX\T \bar{\bz} + \operatorname{tr} \big[(\bG\Had \bOmega)\,\bC_{\bbeta}\big]\Big\}, \label{eq:elbo-lik}\\
    \E_{q}[\log p(\bbeta)] &= -\tfrac{p}{2}\log(2\pi\nu^{2})\;-\;\tfrac{1}{2\nu^{2}}\Big\{\operatorname{tr}(\bSigma)+\bmu\T\bmu\Big\}, \label{eq:elbo-beta}\\
    \E_{q}[\log p(\bgamma)] &= \sum_{j=1}^{p} \Big[ w_{j}\,\log \rho + (1-w_{j})\,\log(1-\rho)\Big].
    \label{eq:elbo-gamma}
\end{align}
Moreover,
\begin{align}
    \E_{q}[\log q(\bbeta)] &= - \tfrac{p}{2}\log(2\pi) - \tfrac{1}{2} \log\det\bSigma - \tfrac{p}{2}, \label{eq:elbo-qbeta}\\
    \E_{q}[\log q(\bz)] &= - \tfrac{n}{2}\log(2\pi) - \tfrac{1}{2}\,\sum_{i=1}^{n} \E_{q(z_{i})}\big[(z_{i}-m_{i})^{2}\big] - \sum_{i=1}^{n}\log\Phi(k_{i} m_{i}), \label{eq:elbo-qz}\\
    \E_{q}[\log q(\bgamma)] &= \sum_{j=1}^{p} \Big[ w_{j}\log w_{j} + (1-w_{j})\log(1-w_{j})\Big], \label{eq:elbo-qgamma}
\end{align}
with $\E_{q(z_{i})}[(z_{i}-m_{i})^{2}]=1-k_{i} m_{i}\,\lambda(k_{i} m_{i})$.
Summing \eqref{eq:elbo-lik} to \eqref{eq:elbo-gamma} and then subtracting the terms \eqref{eq:elbo-qbeta} to \eqref{eq:elbo-qgamma} yields $\mathcal{L}$.

The expression appearing in \eqref{eq:elbo-lik} is obtained by exploiting the following trick for the likelihood term.
Using $\bC_{\bbeta}=\E_{q(\bbeta)}[\bbeta\bbeta\T]=\bSigma+\bmu\bmu\T$, $\E_{q(\bgamma)}[\bGamma \bG\bGamma]=\bG\Had\bOmega$, and recalling $\bbeta\T \bA \bbeta = \operatorname{tr}(\bA \bbeta\bbeta\T)$ (quadratic to trace trick) and $\E \left[\bbeta\T\bA \bbeta \right]= \E \left[\operatorname{tr}(\bbeta\T\bA \bbeta) \right] = \E \left[\operatorname{tr}(\bA \bbeta\bbeta\T) \right] = \operatorname{tr}(\E \left[ \bA \bbeta\bbeta\T\right])$ by linearity of the trace operator, we have
\begin{align*}
    \E_{q}\big[ (\bz - \bX\bGamma\bbeta)\T (\bz - \bX\bGamma\bbeta) \big] &= \sum_{i=1}^{n} \E_{q(z_{i})}[z_{i}^{2}] - 2\,\E_{q(\bbeta)q(\bgamma)}[\bbeta\T \bGamma \bX\T \bz] + \E_{q(\bbeta)q(\bgamma)}[\bbeta\T (\bGamma \bG \bGamma)\bbeta]\\
    &= S_{zz} - 2\, \bmu\T \bW \bX\T \bar{\bz} + \operatorname{tr}\left(\E_{q(\bbeta)q(\bgamma)} \left[ (\bGamma \bG \bGamma) \bbeta\bbeta\T\right] \right)\\
    &= S_{zz} - 2\, \bmu\T \bW \bX\T \bar{\bz} + \operatorname{tr}(\E_{q(\bgamma)} \left[ \bGamma \bG \bGamma \right] \E_{q(\bbeta)} \left[\bbeta\bbeta\T\right])\\
    &= S_{zz} - 2\, \bmu\T \bW \bX\T \bar{\bz} + \operatorname{tr} \big[(\bG\Had \bOmega)\,\bC_{\bbeta}\big],
\end{align*}
which leads to \eqref{eq:elbo-lik}.

\section{Efficient implementation of the MFVB updates}\label{app:impl}
This section summarizes three implementation details used to speed up Algorithm~\ref{algo:1} in high-dimensional regimes: (i) reducing the $\bbeta$-update to the subset of ``active'' predictors, i.e.\ those with non-negligible inclusion probabilities; (ii) applying Woodbury's identity to evaluate $(\bmu,\bSigma)$ when the active dimension is large relative to $n$; and (iii) an optional fast (approximate) update of the inclusion probabilities $\bw$ that avoids $\mathcal{O}(p^{2})$ operations.
These implementation choices are primarily computational: the Woodbury formula provides an algebraically equivalent evaluation of the Gaussian update on the active set, while the active-set restriction itself amounts to thresholding, with coordinates satisfying $w_{j}\le\varepsilon$ treated as inactive, as if $w_{j}=0$.
This is exact when the corresponding weights degenerate to zero, while it is an accurate approximation for sufficiently small $\varepsilon$, e.g.\ $\varepsilon=10^{-10}$.
In our experiments, we took $\varepsilon=0$ to avoid approximations arising from this step, while still reducing the actual dimension of the update in case some weights degenerate to zero.
Finally, the fast $\bgamma$ update introduces an explicit approximation in the $\bgamma$-step (Section~\ref{app:fastgamma}) to avoid $\mathcal{O}(p^{2})$ operations.

\subsection{Active-set reduction for the Gaussian factor}\label{app:active}

Let $\bw=(w_{1},\dots,w_{p})\T$ denote the current inclusion probabilities and fix a numerical threshold $\varepsilon\ge0$.
Define the set of possibly active coordinates (i.e.\ those whose weights did not numerically degenerate to zero) $S_{\bw}=\{j:\,w_{j}>\varepsilon\}$ with $k=|S_{\bw}|$, and let $S_{\bw}^{c}=\{1,\dots,p\}\setminus S_{\bw}$.
Write the coefficient vector as $\bbeta =(\bbeta_{S_{\bw}}\T,\bbeta_{S_{\bw}^{c}}\T)\T$ and partition the design matrix accordingly as $\bX=(\bX_{:S_{\bw}},\bX_{:S_{\bw}^{c}})$.
Here and in the following, for a vector we use the subscript $S_{\bw}$ to denote restriction to the corresponding indices, while for a square matrix it denotes the principal submatrix indexed by $S_{\bw}$.
For a generic matrix $\bX$, we use $\bX_{:S_{\bw}}$ to denote the submatrix formed by the columns indexed by $S_{\bw}$.
In the active-set approximation, we treat the inactive indicators as exactly zero, i.e.\ we set $w_{j}=0$ for $j\in S_{\bw}^{c}$ (equivalently, $\gamma_{j}\equiv 0$ in the linear predictor).
Then, according to this approximation, $\bX\bGamma\bbeta\approx\bX_{:S_{\bw}}\bGamma_{S_{\bw}}\bbeta_{S_{\bw}}$ and the log-likelihood depends only on $\bbeta_{S_{\bw}}$:
\begin{equation*}
  \begin{split}
      \log p(\bz\mid\bbeta,\bgamma) &\approx \log p(\bz\mid\bbeta_{S_{\bw}},\bgamma_{S_{\bw}}) =-\tfrac{1}{2}(\bz - \bX_{:S_{\bw}}\bGamma_{S_{\bw}}\bbeta_{S_{\bw}})\T(\bz - \bX_{:S_{\bw}}\bGamma_{S_{\bw}}\bbeta_{S_{\bw}})\\
      &= -\tfrac{1}{2} \bbeta_{S_{\bw}}\T \bGamma_{S_{\bw}} \bG_{S_{\bw}} \bGamma_{S_{\bw}}\bbeta_{S_{\bw}} + \bbeta_{S_{\bw}}\T \bGamma_{S_{\bw}} \bX_{:S_{\bw}}\T \bz + \text{const},
  \end{split}
\end{equation*}
with $\bG_{S_{\bw}}=\bX_{:S_{\bw}}\T\bX_{:S_{\bw}}$.
Exploiting the prior independence assumption, we can write $p(\bbeta)=p(\bbeta_{S_{\bw}})\,p(\bbeta_{S_{\bw}^{c}})$ with $p(\bbeta_{S_{\bw}})=\phi_{k}(\bbeta_{S_{\bw}}; \nu^{2}\bI_{k})$ and $p(\bbeta_{S_{\bw}^{c}})=\phi_{p-k}(\bbeta_{S_{\bw}^{c}}; \nu^{2}\bI_{p-k})$, obtaining
\[
    p(\bbeta\mid\bz,\bgamma)\;\propto\;p(\bz\mid\bbeta_{S_{\bw}},\bgamma_{S_{\bw}})\,p(\bbeta_{S_{\bw}})\,p(\bbeta_{S_{\bw}^{c}}).
\]
Consequently, one gets
\begin{equation*}
    \begin{split}
    \log q(\bbeta) &= \E_{q(\bz)q(\bgamma)} \big[\log p(\bbeta\mid \bz,\bgamma,\by) \big] + \text{const}\\ 
    &= \E_{q(\bz)q(\bgamma_{S_{\bw}})} \big[\log p(\bz\mid\bbeta_{S_{\bw}},\bgamma_{S_{\bw}}) + \log p(\bbeta_{S_{\bw}}) \big] + \log p(\bbeta_{S_{\bw}^{c}}) + \text{const}\\
    &= \E_{q(\bz)q(\bgamma_{S_{\bw}})} \big[ - \tfrac{1}{2}\bbeta_{S_{\bw}}\T\big(\nu^{-2}\bI_{k}+\bGamma_{S_{\bw}} \bG_{S_{\bw}}\bGamma_{S_{\bw}}\big)\bbeta_{S_{\bw}} + \bbeta_{S_{\bw}}\T \bGamma_{S_{\bw}} \bX_{:S_{\bw}}\T \bz \big] \\
    &\hspace{8cm} + \log p(\bbeta_{S_{\bw}^{c}}) + \text{const}\\
    &= -\tfrac{1}{2} \bbeta_{S_{\bw}}\T\big(\nu^{-2}\bI_{k} + \E_{q(\bgamma_{S_{\bw}})}[\bGamma_{S_{\bw}} \bG_{S_{\bw}} \bGamma_{S_{\bw}}]\big)\bbeta_{S_{\bw}} + \bbeta_{S_{\bw}}\T \E_{q(\bgamma_{S_{\bw}})}[\bGamma_{S_{\bw}}] \bX_{:S_{\bw}}\T \E_{q(\bz)}[\bz]\\
    &\hspace{8cm} + \log p(\bbeta_{S_{\bw}^{c}}) + \text{const},
    \end{split}
\end{equation*}
where the first two summands involve only $\bbeta_{S_{\bw}}$ while the third one involves only $\bbeta_{S_{\bw}^{c}}$.
Within the active-set simplification, this means that the optimal $q(\bbeta)$ factorizes as
\begin{equation}
    \label{eq:fact_qbeta}
    q(\bbeta)=q(\bbeta_{S_{\bw}})\,q(\bbeta_{S_{\bw}^{c}}),
\end{equation}
where $q(\bbeta_{S_{\bw}^{c}})=p(\bbeta_{S_{\bw}^{c}})=\phi_{p-k}(\bbeta_{S_{\bw}^{c}}; \nu^{2}\bI_{p-k})$, so that only $q(\bbeta_{S_{\bw}})$ needs to be updated.
As already stated above, the factorization \eqref{eq:fact_qbeta} holds exactly under hard thresholding ($w_{j}=0$ for $j\in S_{\bw}^{c}$), while it provides an accurate approximation when $w_{j}\le\varepsilon$ is small.
The form of $q(\bbeta_{S_{\bw}})$ is obtained from the equality
\begin{equation*}
\begin{split}
    \log q(\bbeta_{S_{\bw}}) &= -\tfrac{1}{2} \bbeta_{S_{\bw}}\T\big(\nu^{-2}\bI_{k} + \E_{q(\bgamma_{S_{\bw}})}[\bGamma_{S_{\bw}} \bG_{S_{\bw}} \bGamma_{S_{\bw}}]\big)\bbeta_{S_{\bw}}\\
    &\qquad\qquad\qquad+ \bbeta_{S_{\bw}}\T \E_{q(\bgamma_{S_{\bw}})}[\bGamma_{S_{\bw}}] \bX_{:S_{\bw}}\T \E_{q(\bz)}[\bz] + \text{const},
\end{split}
\end{equation*}
from which, adapting the computations in the main manuscript to the restricted model on $S_{\bw}$, one gets $q(\bbeta_{S_{\bw}}) = \phi_{k}(\bbeta_{S_{\bw}} - \bmu_{S_{\bw}};\bSigma_{S_{\bw}})$, with $\bSigma_{S_{\bw}}=(\nu^{-2}\bI_{k} + \bG_{S_{\bw}}\odot\bOmega_{S_{\bw}})^{-1}$ and $\bmu_{S_{\bw}}= \bSigma_{S_{\bw}} \bW_{S_{\bw}} \bX_{:S_{\bw}}\T\bar{\bz}$, where $\bOmega_{S_{\bw}} = \E_{q(\bgamma_{S_{\bw}})}[\bgamma_{S_{\bw}} \bgamma_{S_{\bw}}\T]= \bW_{S_{\bw}}(\bI_{k} - \bW_{S_{\bw}})+ \bw_{S_{\bw}}\bw_{S_{\bw}}^{\T}$, with $\bW_{S_{\bw}}=\diag(\bw_{S_{\bw}})$ and $\bw_{S_{\bw}}$ denoting, as usual, the restriction of $\bw$ to $S_{\bw}$.
Notice that the form of $\bmu_{S_{\bw}}$ and $\bSigma_{S_{\bw}}$ corresponds to the form of the parameters obtained for the full update of $q(\bbeta)$, restricted to $S_{\bw}$, while the inactive coordinates satisfy $\bmu_{S_{\bw}^{c}}=\mathbf 0$ and $\bSigma_{S_{\bw}^{c}}=\nu^{2}\bI_{p-k}$.
This reduces the dominant linear algebra from dimension $p$ to dimension $k$, which is typically much smaller in sparse regimes.
In what follows, we provide efficient formulas to evaluate $(\bmu_{S_{\bw}},\bSigma_{S_{\bw}})$, with a further speedup based on Woodbury's identity when $k>n$.

\subsection{Woodbury evaluation when the active set is large}\label{app:woodbury}

Conditionally on an active set $S_{\bw}$ as in Section~\ref{app:active}, the MFVB update for $q(\bbeta_{S_{\bw}})=\phi_{k}(\bbeta_{S_{\bw}}-\bmu_{S_{\bw}};\bSigma_{S_{\bw}})$ involves the $k\times k$ precision matrix
$$
    \bSigma_{S_{\bw}}^{-1}=\nu^{-2}\bI_{k} + \bG_{S_{\bw}}\odot\bOmega_{S_{\bw}}, 
$$
where $\bG_{S_{\bw}}\odot\bOmega_{S_{\bw}}= \bG_{S_{\bw}}\odot \bw_{S_{\bw}}\bw_{S_{\bw}}\T + \bG_{S_{\bw}}\odot \diag(\bw_{S_{\bw}} - \bw_{S_{\bw}}^{2})$.
Since
$$
    \bG_{S_{\bw}}\odot \bw_{S_{\bw}}\bw_{S_{\bw}}^{\T} = \bX_{\bw}^{\T} \bX_{\bw},
$$
with $\bX_{\bw}=\bX_{:S_{\bw}} \bW_{S_{\bw}}$, and, calling $\bs_{S_{\bw}}= \diag (\bG_{S_{\bw}})$, 
$$
    \bG_{S_{\bw}}\odot \diag(\bw_{S_{\bw}} - \bw_{S_{\bw}}^{2})=\diag(\bs_{S_{\bw}}\odot (\bw_{S_{\bw}} - \bw_{S_{\bw}}^{2})),
$$
one can rewrite $\bSigma_{S_{\bw}}^{-1}$ as
\begin{equation*}
    \bSigma_{S_{\bw}}^{-1}=\bD+\bX_{\bw}\T\bX_{\bw},
\end{equation*}
where $\bD=\nu^{-2}\bI_{k}+\diag(\bs_{S_{\bw}}\odot (\bw_{S_{\bw}} - \bw_{S_{\bw}}^{2}))$ is a diagonal matrix.
When $k>n$ it is thus convenient to use Woodbury's identity:
\begin{equation*}
    \bSigma_{S_{\bw}} = \bD^{-1} - \bD^{-1}\bX_{\bw}\T\Big(\bI_{n}+\bX_{\bw}\bD^{-1}\bX_{\bw}\T\Big)^{-1}\bX_{\bw}\bD^{-1}.
\end{equation*}
Calling $\bM_{n}=\bI_{n}+\bX_{\bw}\bD^{-1}\bX_{\bw}\T \in \R^{n\times n}$ for brevity, the mean update $\bmu_{S_{\bw}}= \bSigma_{S_{\bw}} \bW_{S_{\bw}} \bX_{:S_{\bw}}\T\bar{\bz}$ can then be evaluated without forming $\bSigma_{S_{\bw}}$ explicitly by
\begin{equation}
\label{eq:woodbury-mu}
\begin{split}
    \bmu_{S_{\bw}} &= \bSigma_{S_{\bw}} \bW_{S_{\bw}} \bX_{:S_{\bw}}\T\bar{\bz}=
    \bD^{-1}\bX_{\bw}\T\bar\bz 
    -\bD^{-1}\bX_{\bw}\T \bM_{n}^{-1}\bX_{\bw}\bD^{-1}\bX_{\bw}\T\bar\bz\\
    &= \bD^{-1}\bX_{\bw}\T\left[\bI_{n} - \bM_{n}^{-1}\bX_{\bw}\bD^{-1}\bX_{\bw}\T\right]\bar\bz\\
    &= \bD^{-1}\bX_{\bw}\T\left\{\bM_{n}^{-1}\left[\bM_{n} - \bX_{\bw}\bD^{-1}\bX_{\bw}\T \right]\right\}\bar\bz\\
    &=\bD^{-1}\bX_{\bw}\T \bM_{n}^{-1}\bar{\bz}.
\end{split}
\end{equation}
Moreover, notice that the diagonal of $\bSigma_{S_{\bw}}$ (needed for the fast approximate update presented in Section~\ref{app:fastgamma}, as well as for ELBO monitoring) can be obtained efficiently as
\begin{equation}
\label{eq:woodbury-diag}
\diag(\bSigma_{S_{\bw}}) = \diag(\bD^{-1})-\diag\!\Big(\bD^{-1}\bX_{\bw}\T\,\bM_{n}^{-1}\bX_{\bw}\bD^{-1}\Big).
\end{equation}
Finally, the matrix determinant lemma yields
\begin{equation}
\label{eq:woodbury-logdet}
\begin{split}
   \log\det(\bSigma_{S_{\bw}}) &= -\log\det(\bD+\bX_{\bw}\T\bX_{\bw})\\
    &= -\log\det(\bD)-\log\det(\bM_{n}) = -\sum_{j=1}^{k} \log d_{j} - \log\det(\bM_{n}), 
\end{split}
\end{equation}
where $d_{j}$ denotes the $j$th diagonal entry of $\bD$.
Equations~\eqref{eq:woodbury-mu}--\eqref{eq:woodbury-logdet} imply that when $k>n$ the dominant linear algebra involves an $n\times n$ matrix rather than a $k\times k$ matrix.

\subsection{Optional fast update for the inclusion probabilities}\label{app:fastgamma}

The exact coordinate update of $q(\gamma_{j})$ in Section~\ref{subsec:2.3} requires, for each $j$, evaluating the quantity
$\sum_{k\neq j}(\Sigma_{jk}+\mu_{j}\mu_{k})w_{k} G_{jk}$, which can lead to $\mathcal{O}(p^{2})$ work per sweep.
In high-dimensional settings, this might be prohibitive.
Thus, we also consider an optional approximation that ignores the off-diagonal covariances of $q(\bbeta)$ \emph{within the $\bw$-update}, i.e.\ it replaces $\Sigma_{jk}$ by $0$ for $k\neq j$ only when forming $\eta_{j}$.
This yields the approximate logit-scale parameter
\begin{equation*}
    \eta_{j}^{\mathrm{fast}} = \logit(\rho) -\frac12(\Sigma_{jj}+\mu_{j}^{2})\,G_{jj} +\mu_{j}\,\bX_{:j}\T \bar{\bz} -\mu_{j}\sum_{k\neq j}\mu_{k} w_{k} G_{jk},
\end{equation*}
where the last term can be rewritten as
\begin{equation*}
    \begin{split}
        \sum_{k\neq j}\mu_{k} w_{k} G_{jk}
        &=\sum_{k\neq j} \bX_{:j}\T (\mu_{k} w_{k} \bX_{:k}) \\ &=\bX_{:j}\T\Big(\bX\bW\bmu-\bX_{:j}(w_{j}\mu_{j})\Big).
    \end{split}
\end{equation*}
We can thus write
\begin{equation*}
\begin{split}
    \mu_{j}\,\bX_{:j}\T \bar{\bz} -\mu_{j}\sum_{k\neq j}\mu_{k} w_{k} G_{jk} & = \mu_{j}\,\bX_{:j}\T[\bar\bz - \bX\bW\bmu + \bX_{:j}(w_{j}\mu_{j})]\\
    &=\mu_{j}\,\bX_{:j}\T[\br + \bX_{:j}(w_{j}\mu_{j})],
\end{split}
\end{equation*}
where we defined the residual $\br = \bar{\bz}-\bX\bW\bmu$, so that each coordinate update uses $\bX_{:j}\T(\br+\bX_{:j}w_{j}\mu_{j})$.
For each $j$, after updating $w_{j}$, the residual is then updated online at cost $\mathcal{O}(n)$ by the rank-one correction $\br^{new}=\br - (w_{j}^{new}-w_{j}^{old})\mu_{j}\,\bX_{:j}$, recalling that in the $\bgamma$-update $\bar{\bz}$ and $\bmu$ remain constant.
This reduces the cost of a full $\bw$-sweep to $\mathcal{O}(np)$ operations (excluding precomputations).
We stress that this approximation only affects the $\bgamma$-step; the update of $q(\bbeta)$ (and its Woodbury implementation) remains unchanged.
In our experiments, this fast update yields substantial runtime improvements with little impact on predictive deviance and the highest-ranked variables.

\section{Gibbs sampler}
\label{sec:collapsed-pseudocode}
We summarize here the key steps of the blocked Gibbs sampler used in the paper.
For existing MCMC strategies in Bayesian probit models without and with variable selection, we refer, for instance, to \cite{albert1993bayesian, chang2016bayesian}, respectively.

The form of the full conditionals of $\bbeta,\bz,\bgamma$ is available in closed form, as shown in the manuscript.
It is convenient to rewrite the full conditional for $\bbeta$ by isolating the coefficients associated with the active and inactive covariates.
Calling $S = \{j\in \{ 1,\ldots,p \} \colon \gamma_{j}=1\}$ the set of active covariates, one can partition $\bbeta$ into $\bbeta_{S}=(\beta_{j}\colon\ j\in S)\T\in \R^{|S|}$ and $\bbeta_{S^{c}}=(\beta_{j}\colon\ j \in S^{c})\T \in \R^{p-|S|}$, where $S^{c} = \{j\in \{1,\ldots,p \} \colon \gamma_{j}=0\}$ denotes the set of inactive covariates.
Partitioning the vector $\bzeta=\bX\T \bz$ in the same way, one gets
\begin{equation*}
    \log p(\bbeta\mid \bz,\bgamma) = -\tfrac{1}{2 \nu^{2}} \bbeta_{S^{c}}\T\bbeta_{S^{c}} -\tfrac{1}{2} \bbeta_{S}\T\big(\nu^{-2}\bI_{|S|}+\bG_{S}\big)\bbeta_{S} + \bbeta_{S}\T \bzeta_{S} + \text{const},
\end{equation*}
where $\bG_{S}=\bX_{:S}\T \bX_{:S}$ and $\bX_{:S}$ is the $n\times |S|$ matrix obtained by selecting from $\bX$ the columns in $S$.
Thus, calling $\bB_{S}=\nu^{-2}\bI_{|S|}+\bG_{S}$, conditionally on $(\bz,\bgamma)$, one gets
\begin{equation*}
\bbeta_{S^{c}}\sim \N_{p-|S|}(\bm{0},\nu^{2}\bI_{p-|S|}) \indep \bbeta_{S}\sim \N_{|S|}\left(\bB_{S}^{-1} \bzeta_{S},\bB_{S}^{-1} \right),
\end{equation*}
that is, $\bbeta_{S^{c}}$ is distributed according to the prior distribution, while $\bbeta_{S}$ follows the posterior distribution of the reduced model in which only the active variables are considered.
This is particularly useful when implementing MCMC sampling schemes, significantly reducing the dimension of the problem.

To improve mixing and efficiency, $\bgamma$ and $\bbeta$ are updated jointly.
This makes the sampling of the inactive coefficients $\bbeta_{S^{c}}$ unnecessary, since they will not appear in the linear predictor: we can then set them equal to $0$.
Thus, we update $(\bgamma,\bbeta)$ by first updating $\bgamma$ by sampling each coordinate $\gamma_{j}$ from its collapsed full conditional $p(\gamma_{j}\mid \bgamma_{-j},\bz, \by) = p(\gamma_{j}\mid \bgamma_{-j},\bz)$ and then sampling $\bbeta_{S}$ from the full conditional $p(\bbeta_{S} \mid \bgamma, \bz)$ above.
Finally, $\bz$ is updated by sampling from its full conditional.
Thus, we derive below the full conditional $p(\gamma_{j}\mid \bgamma_{-j}, \bz)$ and then summarize the steps of the sampler in Algorithm~\ref{algo:collapsed-gibbs}.
Conditionally on $(\bgamma_{-j},\bz)$, $\gamma_{j}$ is a Bernoulli random variable with probability of success given by
\begin{equation*}
    \Pr[\gamma_{j}=1\mid \bgamma_{-j}, \bz]=
    \dfrac{p(\bz\mid\bgamma_{-j},\gamma_{j}=1) \Pr[\gamma_{j}=1]}{p(\bz\mid\bgamma_{-j},\gamma_{j}=1) \Pr[\gamma_{j}=1] + p(\bz\mid\bgamma_{-j},\gamma_{j}=0) \Pr[\gamma_{j}=0]},
\end{equation*}
so that
\begin{equation*}
    \logit\{ \Pr[\gamma_{j}=1\mid \bgamma_{-j}, \bz]\}=
    \log p(\bz\mid\bgamma_{-j},\gamma_{j}=1) - \log p(\bz\mid\bgamma_{-j}, \gamma_{j}=0) + \logit(\rho).
\end{equation*}
Thus, a key quantity is $\log p(\bz\mid\bgamma)$.
This can be computed easily by recalling that $\bz=\bX\bGamma\bbeta + \beps = \bX_{:S}\bbeta_{S} + \beps$, where $\beps\sim\N_{n}(\bm{0}, \bI_{n})$ and $\bX_{:S}$ is obtained by selecting from $\bX$ the columns whose indices are in $S$.
Thus, marginalizing out $\bbeta_{S}$, we obtain $\bz\mid \bgamma\sim \N_{n}(\bm{0},\bLambda(S))$, where $\bLambda(S)=\bI_{n}+\nu^{2}\bX_{:S} \bX_{:S}\T$.
We can avoid working with $n$-dimensional matrices by exploiting the following properties.
Using the matrix determinant lemma, one has
\begin{equation*}
    \begin{split}
        \det \bLambda(S) &= \det(\bI_{|S|}+\nu^{2}\bX_{:S}\T \bX_{:S})\det(\bI_{n})\\
        &= \det( (\nu^{2} \bI_{|S|})\cdot \bB_{S} )\\
        &= \nu^{2|S|} \det(\bB_{S}).
    \end{split}
\end{equation*}
Exploiting Woodbury's identity, we can also write $\bLambda(S)^{-1}=\bI_{n}-\bX_{:S}\bB_{S}^{-1}\bX_{:S}\T$, so that, recalling that $\bzeta=\bX\T\bz$, it holds
\begin{equation*}
    \log p(\bz\mid \bgamma) = -\tfrac{1}{2}\big(|S|\log\nu^{2}+\log\det\bB_{S}\big)+\tfrac{1}{2}\,\bzeta_{S}\T\bB_{S}^{-1}\bzeta_{S} + \text{const},
\end{equation*}
where the constant term is independent of $\bgamma$.

In the Gibbs update for $\gamma_{j}$, we need $\log p(\bz\mid\bgamma_{-j}, \gamma_{j}=1)$ and $\log p(\bz\mid\bgamma_{-j}, \gamma_{j}=0)$ while keeping $\bgamma_{-j}$ at its most recent values.
To avoid recomputing both quantities at every coordinate, we store the current log-marginal and update it online.

Let $S=\{k:\gamma_{k}=1\}$ as above and write the log-marginal as $L(S)=\log p(\bz\mid S)$ for notational simplicity.
At the start of iteration $t$, compute once
\[
  L \leftarrow L \big(S^{(t-1)}\big) = \log p \big(\bz^{(t-1)} \mid \bgamma^{(t-1)}\big).
\]
Then update $\gamma_{j}$, $j=1,\dots,p$, as follows.
For each $j$, define
\[
  S_{+} = S\cup\{j\}, \qquad S_{-}=S\setminus\{j\}.
\]
There are two cases:
\\
\textbf{Case 1:} $\gamma_{j}^{(t-1)}=0$ ($j\notin S$).
The current $L$ already equals
\[
  L = L(S) = \log p(\bz\mid \bgamma_{-j}, \gamma_{j}=0).
\]
Compute only the ``addition'' value
\[
  L_{+} = L(S_{+}) = \log p(\bz\mid \bgamma_{-j}, \gamma_{j}=1).
\]
Sample $\gamma_{j}^{(t)}$ using $L$ and $L_{+}$.
If $\gamma_{j}^{(t)}=0$, leave $L$ unchanged; if $\gamma_{j}^{(t)}=1$, update the cache and the active set:
\[
  L \leftarrow L_{+}, \qquad S \leftarrow S_{+}.
\]
\textbf{Case 2:} $\gamma_{j}^{(t-1)}=1$ ($j\in S$).
The current $L$ equals
\[
  L = L(S) = \log p(\bz\mid \bgamma_{-j}, \gamma_{j}=1).
\]
Compute only the ``deletion'' value
\[
  L_{-} = L(S_{-}) = \log p(\bz\mid \bgamma_{-j}, \gamma_{j}=0).
\]
Sample $\gamma_{j}^{(t)}$.
If $\gamma_{j}^{(t)}=1$, keep $L$ unchanged; if
$\gamma_{j}^{(t)}=0$, update:
\[
  L \leftarrow L_{-}, \qquad S \leftarrow S_{-}.
\]
Thus, at each coordinate, we evaluate only the alternative state's log-marginal, while the current state's value is the stored $L$.
When a flip occurs, both $S$ and $L$ are updated to stay up to date.
See Algorithm~\ref{algo:collapsed-gibbs} for the full procedure.

\begin{spacing}{1}
\begin{algorithm}[H]
\caption{Blocked Gibbs sampler for sparse probit regression}
\begin{algorithmic}[1]
    \State \textbf{Input:} $\bX\in\R^{n\times p},\,\by\in\{0,1\}^{n}$, $\nu^{2}$, $\rho$; iterations $T$, burn-in $B$.
    \State \textbf{Precompute:} $\bG \gets \bX\T \bX$;\quad define $\log p(\bz\mid S)$ via $\bB_{S}=\nu^{-2}\bI_{|S|}+\bG_{S}$ and $\bzeta=\bX\T\bz$:\\
    $
    \quad \log p(\bz\mid S) \gets -\tfrac{1}{2}\big(|S|\log \nu^{2} + \log\det \bB_{S}\big) + \tfrac{1}{2}\, \bzeta_{S}\T \bB_{S}^{-1}\bzeta_{S}.
    $
    \State \textbf{Initialize} $\bgamma^{(0)}$ and $\bbeta^{(0)}$; draw $\bz^{(0)}$ from $p(\bz\mid\bgamma^{(0)},\bbeta^{(0)}, \by)$.
    \For{$t=1,\dots,T$}
    \State $\bzeta \gets \bX\T \bz^{(t-1)}$;\quad $S \gets \{j:\gamma^{(t-1)}_{j}=1\}$;\quad $L \gets \log p(\bz^{(t-1)}\mid S)$
    \State \textbf{($\bgamma\mid\bz$)} \For{$j=1,\dots,p$}
        \If{$\gamma^{(t-1)}_{j}=1$}
        \State $S_{-}=S\setminus\{j\}$;\quad $L_{-} \gets \log p(\bz^{(t-1)}\mid S_{-})$
        \State $\ell_{1} \gets L + \log\rho$;\quad $\ell_{0} \gets L_{-} + \log(1-\rho)$;\quad $r_{j}\gets \ell_{1}-\ell_{0}$
        \State Draw $\gamma^{(t)}_{j}\sim \Bern(\expit(r_{j}))$
        \If{$\gamma^{(t)}_{j}=0$} \State $S \gets S_{-}$;\ $L\gets L_{-}$ \EndIf
        \Else
        \State $S_{+}=S \cup \{j\}$;\quad $L_{+} \gets \log p(\bz^{(t-1)}\mid S_{+})$
        \State $\ell_{0} \gets L + \log(1-\rho)$;\quad $\ell_{1} \gets L_{+} + \log\rho$;\quad $r_{j}\gets \ell_{1}-\ell_{0}$
        \State Draw $\gamma^{(t)}_{j}\sim \Bern(\expit(r_{j}))$
        \If{$\gamma^{(t)}_{j}=1$} \State $S \gets S_{+}$;\ $L\gets L_{+}$ \EndIf
        \EndIf
      \EndFor
      \State \textbf{($\bbeta \mid\bz,\bgamma$)} Set $\bbeta^{(t)}_{S^{c}}\gets \bm{0}$ (avoid sampling inactive coefficients).\\
      \hspace*{2.3cm} If $|S|>0$ then form $\bB_{S}$ and draw
      \[
        \bbeta^{(t)}_{S}\sim \N \big(\bB_{S}^{-1} \bzeta_{S},\,\bB_{S}^{-1}\big).
      \]
      \State \textbf{($\bz\mid\bgamma,\bbeta,\by$)}
      For $i=1,\dots,n$, draw $z^{(t)}_{i}\sim \N(\bx_{i}\T \bGamma^{(t)}\bbeta^{(t)},1)$ truncated \hspace*{2.8cm} to $ (0,\infty)\ \text{if } y_{i}=1,\ \text{and to } (-\infty,0]\ \text{if } y_{i}=0$, where \hspace*{2.8cm} $\bGamma^{(t)}=\diag(\bgamma^{(t)})$.
      \State \textbf{Store:} If $t>B$, save $\bgamma^{(t)}$, $\bbeta^{(t)}$.
    \EndFor
    \State \textbf{Return:} draws of $\bgamma$ and $\bbeta$.
\end{algorithmic}
\label{algo:collapsed-gibbs}
\end{algorithm}
\end{spacing}

\FloatBarrier

\section{Additional results}\label{app:additionalResults}
In this section, we report additional results for the applications in the manuscript, including the trace plot of the number of active variables for the MCMC in the LSVT Voice Rehabilitation application and a sensitivity analysis of the MFVB approach to different values of $\nu_{0}^{2}$ for both applications.

\FloatBarrier
\subsection{Trace plot of the number of active covariates}
\begin{figure}[H]
    \centering
    \includegraphics[width=\linewidth]{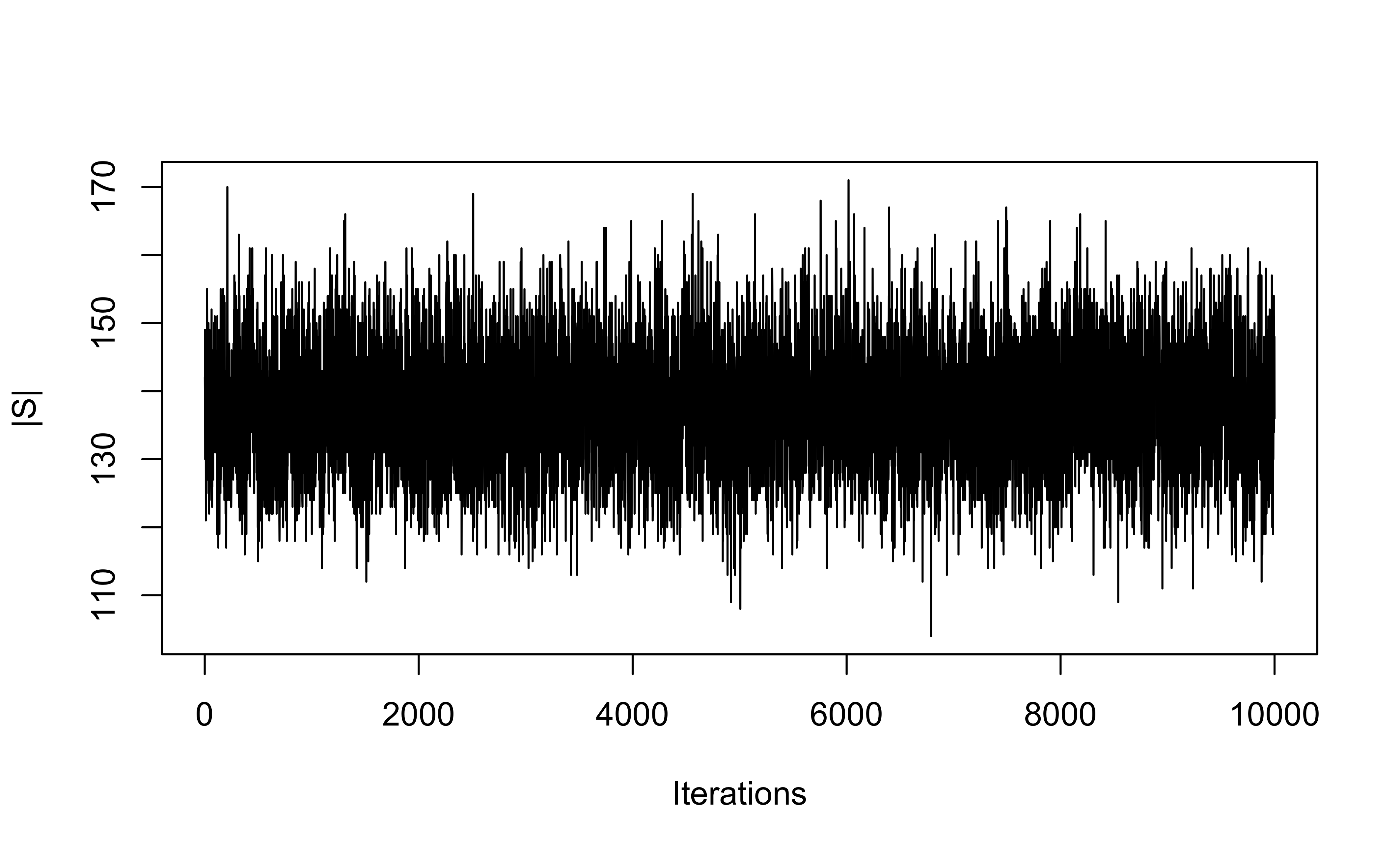}
    \caption{For the LSVT Voice Rehabilitation application, trace plot of the number of active covariates $|S|$ across the $10,000$ MCMC iterations (after burn-in).}
    \label{fig:traceplotVoice}
\end{figure}

We investigate the convergence of the MCMC algorithm for the LSVT Voice Rehabilitation application by inspecting the trace plot of the number of active covariates, namely $|S|$, where $S=\{j:\gamma_{j}=1\}$.
Figure~\ref{fig:traceplotVoice} shows no evidence of non-convergence.

\subsection[Sensitivity analysis for the choice of nu02]{Sensitivity analysis for the choice of $\nu_{0}^{2}$}
For both applications considered in the manuscript, we study how the selected variables and the out-of-sample mean deviances change for different choices of $\nu_{0}^{2}\in\{1,4,9,16,25\}$.
Starting with the LSVT Voice Rehabilitation dataset, Table~\ref{tab:sensitivityVoiceVariables} shows which variables are included for the considered values of $\nu_{0}^{2}$.
As is clear from the tables, all values of $\nu_{0}^{2}$ tend to agree on the majority of the included predictors, with only small differences.
The case $\nu_{0}^{2}=1$ is instead slightly different, as it is the only value for which the predictor \texttt{entropy\_log\_4\_coef} is included, while \texttt{HNR->HNR\_dB\_Praat\_std} is not.
\begin{table}
    \centering
    \begin{tabular}{lccccc}
    \toprule
    & \multicolumn{5}{c}{$\nu_{0}^{2}$}\\
    \cmidrule(lr){2-6}
    Predictor & 1 & 4 & 9 & 16 & 25\\
    \midrule
    Intercept & \checkmark & \checkmark & \checkmark & \checkmark & \checkmark \\
    entropy\_log\_4\_coef & \checkmark & \texttimes & \texttimes & \texttimes & \texttimes \\
    HNR\verb|->|HNR\_dB\_Praat\_std & \texttimes & \checkmark & \checkmark & \checkmark & \checkmark \\
    IMF\verb|->|NSR\_SEO & \checkmark & \checkmark & \checkmark & \checkmark & \checkmark \\
    MFCC\_0th coef & \checkmark & \checkmark & \checkmark & \checkmark & \checkmark \\
    MFCC\_1st coef & \checkmark & \checkmark & \checkmark & \checkmark & \checkmark \\
    MFCC\_7th coef & \texttimes & \texttimes & \checkmark & \checkmark & \checkmark \\
    MFCC\_12th coef & \texttimes & \texttimes & \texttimes & \texttimes & \checkmark \\
    Shimmer\verb|->|Ampl\_abs0th\_perturb & \checkmark & \checkmark & \checkmark & \checkmark & \checkmark \\
\bottomrule
\end{tabular}
\caption{For the LSVT Voice Rehabilitation application, included predictors by MFVB for different values of $\nu_{0}^{2}$.}
\label{tab:sensitivityVoiceVariables}
\end{table}
In terms of out-of-sample average deviance, obtained by averaging the deviance computed on each fold when fitting the model on the remaining folds, the results reported in Table~\ref{tab:sensitivityVoiceDeviances} show small discrepancies for all values of $\nu_{0}^{2}>1$.
Although beyond the scope of the present contribution, one could devise a calibration approach to jointly optimize $\rho$ and $\nu_{0}^{2}$ based on out-of-sample deviance.
\begin{table}[t]
    \centering
    \begin{tabular}{lccccc}
    \toprule
    & \multicolumn{5}{c}{$\nu_{0}^{2}$}\\
    \cmidrule(lr){2-6}
     & 1 & 4 & 9 & 16 & 25\\
    \midrule
    Average deviance & 24.72 & 20.94 & 18.96 & 17.23 & 18.89 \\
\bottomrule
\end{tabular}
\caption{For the LSVT Voice Rehabilitation application, average out-of-sample deviance computed via cross-validation.}
\label{tab:sensitivityVoiceDeviances}
\end{table}
Similarly, we perform an analogous sensitivity analysis for the Alzheimer application.
First, we study which variables are selected for different values of $\nu_{0}^{2}$, noting that values of $\nu_{0}^{2}$ smaller than $25$ do not include the covariates \texttt{Pancreatic\_polypeptide} and \texttt{VEGF}, while a general consensus on the inclusion of \texttt{Ab\_42} and \texttt{tau} is present.
See Table~\ref{tab:sensitivityAlzheimerVariables} for details.
\begin{table}
    \centering
    \begin{tabular}{lccccc}
    \toprule
    & \multicolumn{5}{c}{$\nu_{0}^{2}$}\\
    \cmidrule(lr){2-6}
    Predictor & 1 & 4 & 9 & 16 & 25\\
    \midrule
    Intercept & \checkmark & \checkmark & \checkmark & \checkmark & \checkmark \\
    Ab\_42 & \texttimes & \checkmark & \checkmark & \checkmark & \checkmark \\
    GRO\_alpha & \texttimes & \texttimes & \texttimes & \checkmark & \checkmark \\
    Pancreatic\_polypeptide & \texttimes & \texttimes & \texttimes & \texttimes & \checkmark \\
    tau & \texttimes & \checkmark & \checkmark & \checkmark & \checkmark \\
    VEGF & \texttimes & \texttimes & \texttimes & \texttimes & \checkmark \\
\bottomrule
\end{tabular}
\caption{For the Alzheimer application, included predictors by MFVB for different values of $\nu_{0}^{2}$.}
\label{tab:sensitivityAlzheimerVariables}
\end{table}
Secondly, we inspect the resulting average out-of-sample deviances, obtained by averaging the deviance computed on each fold when fitting the model on the remaining folds.
The results, reported in Table~\ref{tab:sensitivityAlzheimerDeviances}, show that the considered value $\nu_{0}^{2}=25$ corresponds to the best deviance value, while the worst performance is obtained, as expected, from the model corresponding to $\nu_{0}^{2}=1$, which selects only the intercept.
\begin{table}
    \centering
    \begin{tabular}{lccccc}
    \toprule
    & \multicolumn{5}{c}{$\nu_{0}^{2}$}\\
    \cmidrule(lr){2-6}
     & 1 & 4 & 9 & 16 & 25\\
    \midrule
    Average deviance & 85.63 & 74.23 & 65.68 & 62.32 & 60.53 \\
\bottomrule
\end{tabular}
\caption{For the Alzheimer application, average out-of-sample deviance computed via cross-validation.}
\label{tab:sensitivityAlzheimerDeviances}
\end{table}
\FloatBarrier

\printbibliography
\end{refsection}

\end{document}